\documentclass[nofootinbib,preprint,preprintnumbers,a4paper,10pt]{revtex4}
\usepackage{multirow}
\usepackage{textcomp}
\usepackage{amsmath,graphicx,color,epsfig}

\usepackage{footnote}
\usepackage{ulem}
\usepackage{booktabs}
\usepackage{array}
\usepackage{amssymb}
\usepackage{subfigure}
\usepackage{longtable}
\usepackage{verbatim}
\usepackage{amsfonts}
\usepackage{hyperref}
\usepackage{cancel}

\begin{document}

\title{Topological amplitudes of bottom baryon decays in the $SU(3)_F$ limit}

\author{Di Wang$^{1}$}\email{wangdi@hunnu.edu.cn}
\author{Wei-Chen Fu$^{1}$}
\address{%
$^1$Department of Physics, Hunan Normal University, Changsha 410081, China
}

\begin{abstract}
Motivated by the first observation of CP violation in baryon decays, we study the topological amplitudes of bottom baryon decays in the $SU(3)_F$ limit.
The topological diagrams of the charmless two-body decays of bottom baryons are presented in detail.
The linear relations between topologies and $SU(3)$ irreducible amplitudes are derived through tensor contraction and $SU(3)$ decomposition.
Four amplitudes among the 13 independent amplitudes are critical to the CP asymmetries.
The small CP asymmetries might indicate small relative strong phases between amplitudes $A_2$ and $A_{12,14}^\prime$.
To avoid them, we suggest measuring CP asymmetries in the $\Xi^0_b\to pK^-$ and $\Xi^-_b\to \Lambda^0 K^-$ decays.
Furthermore, the K\"orner-Pati-Woo theorem can be tested by measuring the branching fractions of the $\Lambda_b^0\to\Sigma^0K^0_S$ and $\Lambda_b^0\to\Sigma^-K^+$ modes.

\end{abstract}

\maketitle

\section{Introduction}
Bottom baryon decays play an important role in studying perturbative and nonperturbative strong interactions, extracting the quark-mixing Cabibbo-Kobayashi-Maskawa (CKM) matrix elements, and searching for new physics beyond the Standard Model.
Compared with the bottom meson decays, the non-zero spin of baryons provides opportunities to study decay dynamics via more observables beyond branching fractions \cite{Wang:2024qff,Zhang:2022emj,Zhang:2022iye,Zhao:2024ren,Geng:2021sxe,Gronau:2015gha,Ajaltouni:2004zu,Wang:2022fih,
Rui:2022jff,Durieux:2016nqr}.
As bottom baryons are increasingly generated at the Large Hadron Collider (LHC), several charmless decay channels of bottom baryons have been observed in experiments \cite{LHCb:2024iis,LHCb:2016yco,LHCb:2018fpt,LHCb:2019oke,LHCb:2016rja,LHCb:2015kmm,LHCb:2018fly,LHCb:2024yzj,LHCb:2016hwg,LHCb:2014yin}.
Very recently, the LHCb Collaboration has reported the first observation of CP violation in bottom baryon decays \cite{LHCb:2025ray}
\begin{align}
 A_{CP}(\Lambda_b^0\to pK^-\pi^+\pi^-) = (2.45\pm 0.46\pm0.10)\%.
\end{align}
It is a milestone in particle physics as CP asymmetries are well established in meson systems \cite{LHCb:2019hro,Belle:2001zzw,Christenson:1964fg,BaBar:2001ags}, whereas CP violation in baryon decays has not been observed until now.

From a theoretical perspective, the QCD in baryon non-leptonic decays are complicated owing to the presence of three valence quarks \cite{Wang:2024oyi,Duan:2024zjv,Zhu:2018jet,Han:2024kgz,Lu:2009cm}.
Because of the absence of nonperturbative input, only a few decay channels can be calculated in the QCD inspired approach.
To extract the decay information of bottom baryon, it is significant to analyze the topological diagrams in the flavor $SU(3)$ symmetry.
A topological diagram provides an intuitive description of the dynamics of heavy hadron decays and a theoretical framework that allows not only model-dependent data analysis but also model calculations.
Topological diagrams of charmed baryon decays have been widely analyzed in recent studies \cite{Wang:2024nxb,Cheng:2024lsn,Groote:2021pxt,Hsiao:2020iwc,Zhao:2018mov,He:2018joe,Zhong:2024qqs,Hsiao:2021nsc}.
However, the topological amplitudes of bottom baryon decays have not been systematically studied so far.

Topological diagrams in heavy hadron decays can be formalized as invariant tensors constructed from four-quark operators and initial/final states \cite{He:2018php}, allowing us to study topologies with mathematical tools.
Owing to the different spin wave functions of the $q_1\leftrightarrow q_2$ symmetric and antisymmetric octets $\mathcal{B}_8^S$ and $\mathcal{B}_8^A$, the topological diagrams of baryon decays into octet baryons have two distinct sets.
Furthermore, the relations between topological diagrams and $SU(3)$ irreducible amplitudes are not evident, as the topological amplitudes of the $\mathcal{B}_{b\overline 3}\to \mathcal{B}_8M$ decays are constructed from third-rank octet tensors, whereas $SU(3)$ irreducible amplitudes are constructed from $(1,1)$-rank octet tensors.
In Ref.~\cite{Wang:2024ztg}, we established a framework to analyze the topological amplitudes of charmed baryon decays in the $SU(3)_F$ limit.
In this study, we extend this framework to the bottom baryon decays.
The linear relations between topologies and $SU(3)$ irreducible amplitudes are derived through tensor contraction and $SU(3)$ decomposition.
Among the 13 independent amplitudes contributing to $\mathcal{B}_{b\overline 3}\to \mathcal{B}_8 M$ decays, four amplitudes associated with three-dimensional irreducible representations are critical to the CP asymmetries of bottom baryon decays.
Recent measurements of the CP asymmetries of the $\Lambda^0_b\to p\pi^-$ and $\Lambda^0_b\to pK^-$ modes \cite{LHCb:2024iis} have indicated small strong phases between $A_2$ and $A_{12,14}^\prime$.
To avoid the small strong phases, we suggest measuring CP asymmetries in the $\Xi^0_b\to pK^-$ and $\Xi^-_b\to \Lambda^0 K^-$ decays.
Furthermore, the K\"orner-Pati-Woo theorem \cite{Pati:1970fg,Korner:1970xq} can be tested by measuring the branching fractions of $\Lambda_b^0\to\Sigma^0K^0_S$ and $\Lambda_b^0\to\Sigma^-K^+$ decays under isospin symmetry.

The remainder of this paper is organized as follows.
In Sec.~\ref{sm}, we present the topological amplitudes of bottom baryon decays in the $SU(3)_F$ limit.
The phenomenological analysis of the topological amplitudes is presented in Sec.~\ref{pa}.
Sec.~\ref{summary} provides a brief summary.
The discussions about the $SU(3)$ irreducible amplitudes constructed by the third-rank octet tensors are discussed in Appendix.~\ref{su3}.

\section{Topological amplitudes of bottom baryon decays}\label{sm}

The effective Hamiltonian of the $b\to u\overline u q$ transition is given by \cite{Buchalla:1995vs}
\begin{align}\label{hsmb2}
 \mathcal H_{\rm eff}=&{\frac{G_F}{\sqrt 2} }
 \sum_{q=d,s}\left[V_{ub}V_{uq}^*\left(\sum_{i=1}^2C_i^u(\mu)O_i^u(\mu)\right) + V_{cb}V_{cq}^*\left(\sum_{i=1}^2C_i^c(\mu)O_i^c(\mu)\right)\right]\nonumber\\&
 -{\frac{G_F}{\sqrt 2}}\sum_{q=d,s}\left[V_{tb}V_{tq}^*\left(\sum_{i=3}^{10}C_i(\mu)O_i(\mu)
 +C_{7\gamma}(\mu)O_{7\gamma}(\mu)+C_{8g}(\mu)O_{8g}(\mu)\right)\right]+h.c..
 \end{align}
The magnetic-penguin contributions can be included in the Wilson coefficients for the penguin operators
\cite{Beneke:2003zv,Beneke:2000ry,Beneke:1999br}.
In the flavor $SU(3)$ limit, the weak Hamiltonian of bottom decay can be written as \cite{Wang:2020gmn}
 \begin{equation}\label{h}
 \mathcal H_{\rm eff}= \sum_{i,j,k=1}^3 \{H^{(u)k}_{ij}\mathcal{O}_{ij}^{(u)k}+H^{(c)}_{i}\mathcal{O}_{i}^{(c)}
 +H^{(p)k}_{ij}\mathcal{O}_{ij}^{(p)k}\},
 \end{equation}
where $\mathcal{O}_{ij}^{(u)k}$, $\mathcal{O}_{i}^{(c)}$ and $\mathcal{O}_{ij}^{(p)k}$ denote the four-quark operators including the Fermi coupling constant $G_F$ and the Wilson coefficients.
The superscripts $u$, $c$, and $p$ are used to distinguish the tree operators $\mathcal{O}_{1,2}^{(u)}$, $\mathcal{O}_{1,2}^{(c)}$ and the penguin operators $\mathcal{O}_{3-10}$.
Indices $i$, $j$, and $k$ are flavor indices.
The color indices and current structures of the four quark operators are summed into $\mathcal{O}_{ij}^{(u)k}$, $\mathcal{O}_{i}^{(c)}$, and $\mathcal{O}_{ij}^{(p)k}$.
The matrices $H^{(u,c,p)}$ are the coefficient matrices.
According to the effective Hamiltonian of $b\to u\overline uq$ decays, the non-zero CKM coefficients include
\begin{align}
  H^{(u)1}_{21} & = V_{ub}V_{ud}^*,\qquad H^{(u)1}_{31} = V_{ub}V_{us}^*, \qquad H^{(c)}_{2} = V_{cb}V_{cd}^*,\qquad H^{(c)}_{3} = V_{cb}V_{cs}^*, \nonumber\\
 H^{(p)1}_{12} & = H^{(p)2}_{22}=H^{(p)3}_{32}=-V_{tb}V_{td}^*, \qquad H^{(p)1}_{13} = H^{(p)2}_{23}=H^{(p)3}_{33}= -V_{tb}V_{ts}^*.
\end{align}
Similar to the effective Hamiltonian, the initial and final states, for instance $M$, can be written as
\begin{align}
  |M^\alpha\rangle = (M^\alpha)^{i}_{j}|M^{i}_{j} \rangle,
\end{align}
where $|M^{i}_{j} \rangle$ is the quark composition of the meson state, $|M^{i}_{j} \rangle = |q_i\bar q_j\rangle$, and $(M^\alpha)$ is the coefficient matrix.

The decay amplitude of the $\mathcal{B}^\gamma_{b\overline 3}\to \mathcal{B}^\alpha_{8} M^\beta$ mode is constructed as follows:
\begin{align}\label{amp}
\mathcal{A}(\mathcal{B}^\gamma_{b\overline 3}\to \mathcal{B}_{8}^\alpha M^\beta)& = \langle \mathcal{B}_{8}^\alpha M^\beta |\mathcal{H}_{\rm eff}| \mathcal{B}^\gamma_{b\overline 3}\rangle\nonumber\\&~= \sum_{\omega}\,(\mathcal{B}_{8}^\alpha)^{ijk}\langle \mathcal{B}_{8}^{ijk}|(M^\beta)^l_m\langle M^l_m||H_{np}^q\mathcal{O}_{np}^q||(\mathcal{B}^\gamma_{b\overline 3})_{rs}|[\mathcal{B}_{b\overline 3}]_{rs}\rangle\nonumber\\& ~~= \sum_{ \omega}\,\langle \mathcal{B}_{8}^{ijk} M^l_m |\mathcal{O}_{np}^q|[\mathcal{B}_{b\overline 3}]_{rs}\rangle \times (\mathcal{B}_{8}^\alpha)^{ijk}(M^\beta)^l_m H_{np}^q(\mathcal{B}^\gamma_{b\overline 3})_{rs}\nonumber\\&~~~~
= \sum_\omega X_{\omega}(C_\omega)_{\alpha\beta\gamma}.
\end{align}
In the above formula, $\sum_{\omega}$ represents summing over all possible full contractions.
$X_\omega = \langle \mathcal{B}_{8}^{ijk} M^l_m |\mathcal{O}_{np}^q|[\mathcal{B}_{b\overline 3}]_{rs}\rangle$ is the reduced matrix element and $(C_\omega)_{\alpha\beta\gamma}=(\mathcal{B}_{8}^\alpha)^{ijk}(M^\beta)^l_m H_{np}^q(\mathcal{B}^\gamma_{b\overline 3})_{rs}$ is the Clebsch-Gordan (CG) coefficient.
According to the Wigner-Eckart theorem \cite{Eckart30,Wigner59}, $X_\omega$ is independent of the indices $\alpha$, $\beta$, and $\gamma$.
All the information about the initial and final states is absorbed into the Clebsch-Gordan coefficient $(C_\omega)_{\alpha\beta\gamma}$.
There are two different octets with symmetric and antisymmetric flavor wavefunctions under $q_1\leftrightarrow q_2$.
They are labeled by $\mathcal{B}_8^S$ and $\mathcal{B}_8^A$ and their flavor and spin wave functions are $\phi_S\chi_S$ and $\phi_A\chi_A$, respectively.
The total amplitude for the $\mathcal{B}_{b\overline 3}\to \mathcal{B}_8 M$ decay is obtained by summing the amplitudes of the $\mathcal{B}_{b\overline 3}\to \mathcal{B}_8^S M$ and $\mathcal{B}_{b\overline 3}\to \mathcal{B}_8^A M$ transitions,
\begin{align}\label{a6}
  \mathcal{A}(\mathcal{B}_{b\overline 3}\to \mathcal{B}_8 M) = \mathcal{A}^S(\mathcal{B}_{b\overline 3}\to \mathcal{B}_8^S M) + \mathcal{A}^A(\mathcal{B}_{b\overline 3}\to \mathcal{B}_8^A M).
\end{align}

The topological amplitudes for the $\mathcal{B}_{b\overline 3}\to \mathcal{B}_8 M$ decays are similar to those for the $\mathcal{B}_{c\overline 3}\to \mathcal{B}_8 M$ decays, which have been studied in Ref.~\cite{Wang:2024ztg}.
Hence, the details of the model-independent analysis of topological amplitudes are presented in Appendix.~\ref{su3}.
The difference between the $\mathcal{B}_{b\overline 3}\to \mathcal{B}_8 M$ and $\mathcal{B}_{c\overline 3}\to \mathcal{B}_8 M$ decays is that the $d$- and $s$-quark loop diagrams in the $\mathcal{B}_{c\overline 3}\to \mathcal{B}_8 M$ decays are identical under the $SU(3)_F$ symmetry, whereas the $u$- and $c$-quark loop diagrams in the  $\mathcal{B}_{b\overline 3}\to \mathcal{B}_8 M$ decays are distinct, since $c$ quark is not a basis of the flavor $SU(3)$ group.
Hence, the charm-loop amplitudes induced by operator $\mathcal{O}_{i}^{(c)}$ are presented explicitly in Appendix.~\ref{su3}.

To derive the linear relations between the topological diagrams, in which the baryon octet is written as a tensor with three covariant indices, and the $SU(3)$ irreducible amplitudes, in which the baryon octet is written as a tensor with a covariant index and a contravariant index, the key step is to determine the relation between the two different tensor representations.
As noted in Ref.~\cite{Wang:2024ztg}, the third-rank tensors $(\mathcal{B}^S_8)^{ijk}$ and $(\mathcal{B}^A_8)^{ijk}$ are expressed in terms of the ($1,1$)-rank tensor $(\mathcal{B}_8)^i_j$ as
\begin{eqnarray}\label{sy1}
(\mathcal{B}^S_8)^{ijk} = \frac{1}{\sqrt{6}}\left[\epsilon^{kil}(\mathcal{B}_8)^j_l
+\epsilon^{kjl}(\mathcal{B}_8)^i_l\right],\qquad\quad  (\mathcal{B}^A_8)^{ijk}= \frac{1}{\sqrt{2}}\epsilon^{ijl}(\mathcal{B}_8)^k_l.
\end{eqnarray}
One can verify that the indices $i$ and $j$ are symmetric in $(\mathcal{B}^S_8)^{ijk}$ and antisymmetric in $(\mathcal{B}^A_8)^{ijk}$.
The detailed derivation of the linear relations between the topologies and the $SU(3)$ irreducible amplitudes is shown in Appendix.~\ref{su3}.

%%%%%%%%%%%%%%%%%%%%%%%%%%%%%%%%%%%%%%%%%%%%%%%%%%%%%%%%%%%%%%
\begin{table*}
\caption{Topological amplitudes of $\Xi^{-,0}_b$ decays into $\mathcal{B}_8^SM$ from the $b\to d$ transition.}\label{ta1}
 \scriptsize
\begin{tabular}{|c|c|c|c|}
\hline\hline
 Channel & Amplitude \\\hline
 $\Xi_b^-\to \Sigma^0\pi^-$ & $\frac{1}{2\sqrt{3}}\lambda_c(a^{S,c}_1+a^{S,c}_2-2 a^{S,c}_3-3 a^{S,c}_4)-\frac{1}{2\sqrt{3}}\lambda_t(a^{S,p}_1+3 a^{S,p}_{11}-3 a^{S,p}_{12}-3 a^{S,p}_{13}+a^{S,p}_{18}+a^{S,p}_{19}+a^{S,p}_2-2 a^{S,p}_{20}-3 a^{S,p}_{21}$\\&$+3 a^{S,p}_{23}+3 a^{S,p}_{24}-6 a^{S,p}_{25}-9 a^{S,p}_{26}-2 a^{S,p}_3-a^{S,p}_4-a^{S,p}_5+2 a^{S,p}_6+a^{S,p}_7+a^{S,p}_8-2 a^{S,p}_9)$\\&$+\frac{1}{2\sqrt{3}}\lambda_u(a^{S}_1-3 a^{S}_{14}+a^{S}_{18}+a^{S}_{19}+a^{S}_2-2 a^{S}_{20}-3 a^{S}_{21}-2 a^{S}_3)$  \\\hline
 $\Xi_b^-\to\Sigma^-\pi^0$ & $-\frac{1}{2\sqrt{3}}\lambda_c(a_1^{S,c}+a_2^{S,c}-2a_3^{S,c}-3a_4^{S,c})
 +\frac{1}{2\sqrt{3}}\lambda_t(a_1^{S,p}+3a_{11}^{S,p}-3a_{12}^{S,p}-3a_{13}^{S,p}$
 \\&$+a_{18}^{S,p}+a_{19}^{S,p}
 +a_2^{S,p}-2a_{20}^{S,p}-3a_{21}^{S,p}+3a_{23}^{S,p}+3a_{24}^{S,p}-6a_{25}^{S,p}-9a_{26}^{S,p}-2a_3^{S,p}$\\&$
 -a_4^{S,p}-a_5^{S,p}+2a_6^{S,p}+a_7^{S,p}+a_8^{S,p}-2a_9^{S,p})
 -\frac{1}{2\sqrt{3}}\lambda_u(3a_{13}^{S}+a_{18}^{S}+a_{19}^{S}-2a_{20}^{S}
 -3a_{21}^{S})$   \\\hline
 $\Xi_b^-\to\Sigma^-\eta_8$ & $-\frac{1}{2}\lambda_c(a_1^{S,c}-a_2^{S,c}+a_4^{S,c})
 +\frac{1}{2}\lambda_t(-a_1^{S,p}-a_{11}^{S,p}+a_{12}^{S,p}
 +a_{13}^{S,p}+a_{18}^{S,p}-a_{19}^{S,p}+a_2^{S,p}$\\
\hline
  \hline
\end{tabular}
\end{table*}
%%%%%%%%%%%%%%%%%%%%%%%%%%%%%%%%%%%%%%%%%%%%%%%%%%%%%%%%%%%%%%%%%%%%%%%%%%%
\begin{table*}
\caption{Topological amplitudes of $\Lambda^0_b$ decays into $\mathcal{B}_8^SM$ from the $b\to d$ transition.}\label{ta2}
 \scriptsize
\begin{tabular}{|c|c|c|c|}
\hline\hline
 Channel & Amplitude \\\hline
 $\Lambda_b^0\to\Sigma^0K^0$ & $\frac{1}{2\sqrt{3}}\lambda_c(2a_1^{S,c}-a_2^{S,c}-a_3^{S,c})
 +\frac{1}{2\sqrt{3}}\lambda_t(a_1^{S,p}-2a_{18}^{S,p}+a_{19}^{S,p}
 -2a_2^{S,p}+a_{20}^{S,p}-6a_{23}^{S,p}+3a_{24}^{S,p}
 +3a_{25}^{S,p}+a_3^{S,p}+2a_4^{S,p}$\\&$
 -a_5^{S,p}-a_6^{S,p}
 -2a_7^{S,p}+a_8^{S,p}+a_9^{S,p})+\frac{1}{2\sqrt{3}}\lambda_u(2a_{18}^{S}
 -a_{19}^{S}-a_{20}^{S}-2a_4^{S}+a_5^{S}
 +a_6^{S}-2a_7^{S}+a_8^{S}+a_9^{S})$   \\\hline
 $\Lambda_b^0\to\Sigma^-K^+$ & $\frac{1}{\sqrt{6}}\lambda_c(-2a_1^{S,c}+a_2^{S,c}+a_3^{S,c})
 -\frac{1}{\sqrt{6}}\lambda_t(a_1^{S,p}-2a_{18}^{S,p}+a_{19}-2a_2^{S,p}+a_{20}^{S,p}
 -6a_{23}^{S,p}+3a_{24}^{S,p}+3a_{25}^{S,p}$\\&$
 +a_3^{S,p}+2a_4^{S,p}-a_5^{S,p}-a_6^{S,p}-2a_7^{S,p}
 +a_8^{S,p}+a_9^{S,p})+\frac{1}{\sqrt{6}}\lambda_u(-2a_{10}^{S}
 +a_{11}^{S}+a_{12}^{S}-2a_{18}^{S}+a_{19}^{S}+a_{20}^S)$   \\\hline
 $\Lambda_b^0\to p\pi^-$ & $\frac{1}{\sqrt{6}}\lambda_c(-a_1^{S,c}+2a_2^{S,c}-a_3^{S,c}-3a_4^{S,c})
 +\frac{1}{\sqrt{6}}\lambda_t(-2a_1^{S,p}
 -3a_{11}^{S,p}+3a_{12}^{S,p}+3a_{13}^{S,p}+a_{18}^{S,p}-
 2a_{19}^{S,p}+a_2^{S,p}+a_{20}^{S,p}$\\&$
 +3a_{21}^{S,p}+3a_{23}^{S,p}-6a_{24}^{S,p}+3a_{25}^{S,p}+9a_{26}^{S,p}+a_3^{S,p}-a_4^{S,p}+2a_5^{S,p}
 -a_6^{S,p}+a_7^{S,p}-2a_8^{S,p}+a_9^{S,p})
 $\\&$
 -\frac{1}{\sqrt{6}}\lambda_u((a_1^{S}+3a_{14}^{S}+a_{18}^{S}
 -2a_{19}^{S}-2a_2^{S}+a_{20}^{S}+3a_{21}^{S}+a_3^{S}-a_4^{S}-a_5^{S}+2a_6^{S}-a_7^{S}-a_8^{S}+2a_9^{S})$   \\
\hline
   \hline
\end{tabular}
\end{table*}
%%%%%%%%%%%%%%%%%%%%%%%%%%%%%%%%%%%%%%%%%%%%%%%%%%%%%%%%%%%%%%%
\begin{table*}
\caption{Topological amplitudes of $\Xi^{-,0}_b$ decays into $\mathcal{B}_8^SM$ from the $b\to s$ transition.  }\label{ta3}
 \scriptsize
\begin{tabular}{|c|c|c|c|}
\hline\hline
 Channel & Amplitude \\\hline
 $\Xi_b^-\to\Sigma^0K^-$ & $-\frac{1}{2\sqrt{3}}\lambda_c^\prime (a_1^{S,c}-2a_2^{S,c}+a_3^{S,c}+3a_4^{S,c})-\frac{1}{2\sqrt{3}}
 \lambda_t^\prime(2a_1^{S,p}+3a_{11}^{S,p}
 -3a_{12}^{S,p}-3a_{13}^{S,p}-3a_{18}^{S,p}$\\&$
 +2a_{19}^{S,p}-a_2^{S,p}-a_{20}^{S,p}
 -3a_{21}^{S,p}-3a_{23}^{S,p}+6a_{24}^{S,p}-3a_{25}^{S,p}
 -9a_{26}^{S,p}-a_3^{S,p}+a_4^{S,p}
 -2a_5^{S,p}+a_6^{S,p}
 $\\&$-a_7^{S,p}
 +2a_8^{S,p}-a_9^{S,p})+\frac{1}{2\sqrt{3}}\lambda_{u}^\prime(-a_1^{S}-3a_{14}^{S}-a_{18}^{S}
 +2a_{19}^{S}+2a_2^{S}
 -a_{20}^{S}-3a_{21}^{S}-a_3^{S})$     \\\hline
 $\Xi_b^-\to\Sigma^-\overline{K}^0$ & $-\frac{1}{\sqrt{6}}\lambda_c^\prime(a_1^{S,c}-2a_2^{S,c}+a_3^{S,c}+3a_4^{S,c})+\frac{1}{\sqrt{6}}\lambda_t^\prime(-2 a_1^{S,p}-3 a_{11}^{S,p}+3 a_{12}^{S,p}$\\&$
 +3 a_{13}^{S,p}+a_{18}^{S,p}-2 a_{19}^{S,p}+a_2+a_{20}^{S,p}+3 a_{21}^{S,p}+3 a_{23}^{S,p}-6 a_{24}^{S,p}+3 a_{25}^{S,p}+9 a_{26}^{S,p}+a_3^{S,p}-a_4^{S,p}$\\&$+2 a_5^{S,p}-a_6^{S,p}+a_7^{S,p}-2 a_8^{S,p}+a_9^{S,p})
 +\frac{1}{\sqrt{6}}\lambda_u^\prime(-a_{18}^{S}+2a_{19}^{S}
 -a_{20}^{S}-3a_{21}^{S})$     \\\hline
 $\Xi_b^-\to\Xi^0\pi^-$ & $\frac{1}{\sqrt{6}}\lambda_c^\prime(-2a_1^{S,c}+a_2^{S,c}+a_3^{S,c})
 -\frac{1}{\sqrt{6}}\lambda_t^\prime(a_1^{S,p}-2 a_{18}^{S,p}+a_{19}^{S,p}-2 a_2^{S,p}+a_{20}^{S,p}-6 a_{23}^{S,p}+3 a_{24}^{S,p}+3 a_{25}^{S,p}+a_3^{S,p}$\\
\hline
  \hline
\end{tabular}
\end{table*}
%%%%%%%%%%%%%%%%%%%%%%%%%%%%%%%%%%%%%%%%%%%%%%%%%%%%%%%%%%%%%%%%%%%%%%%%%%%%%%%%%%%
\begin{table*}
\caption{Topological amplitudes of $\Lambda_b^0$ decays into $\mathcal{B}_8^SM$ from the $b\to s$ transition.}\label{ta4}
 \scriptsize
\begin{tabular}{|c|c|c|c|}
\hline\hline
 Channel & Amplitude \\\hline
 $\Lambda_b^0\to\Sigma^+\pi^-$ & $
 \frac{1}{\sqrt{6}}\lambda_c^\prime(a_1^{S,c}-2a_2^{S,c}+a_3^{S,c})
 +\frac{1}{\sqrt{6}}\lambda_t^\prime(2 a_1^{S,p}-a_{18}^{S,p}+2 a_{19}^{S,p}-a_2^{S,p}-a_{20}^{S,p}-3 a_{23}^{S,p}+6 a_{24}^{S,p}-3 a_{25}^{S,p}-a_3^{S,p}+a_4^{S,p}$\\&$
 -2 a_5^{S,p}+a_6^{S,p}-a_7^{S,p}+2 a_8^{S,p}-a_9^{S,p})+\frac{1}{\sqrt{6}}\lambda_u^\prime (a_1^{S}+a_{18}^{S}-2a_{19}^{S}-2a_2^{S}+a_{20}^{S}+a_3^{S}-a_4^{S}-a_5^{S}+2a_6^{S})$    \\\hline
 $\Lambda_b^0\to\Sigma^0\pi^0$ & $\frac{1}{\sqrt{6}}\lambda_c^\prime(a_1^{S,c}-2a_2^{S,c}+a_3^{S,p})
 +\frac{1}{\sqrt{6}}\lambda_t^\prime(2 a_1^{S,p}-a_{18}^{S,p}+2 a_{19}^{S,p}-a_2^{S,p}-a_{20}^{S,p}-3 a_{23}^{S,p}+6 a_{24}^{S,p}-3 a_{25}^{S,p}-a_3^{S,p}+a_4^{S,p}-2 a_5^{S,p}$\\&$
 +a_6^{S,p}-a_7^{S,p}+2 a_8^{S,p}-a_9^{S,p})+\frac{1}{2\sqrt{6}}\lambda_u^\prime(a_1^{S}+a_{10}^{S}
 -2a_{11}^{S}+a_{12}^{S}+2a_{18}^{S}-4a_{19}^{S}
 -2a_2^{S}+2a_{20}^{S}+a_3^{S}-a_4^{S}-a_5^{S}+2a_6^{S})$    \\\hline
 $\Lambda_b^0\to\Sigma^0\eta_8$ & $\frac{1}{6\sqrt{2}}\lambda_u^\prime(a_1^{S}+a_{10}^{S}-2a_{11}^{S}
 +a_{12}^{S}-2a_2^{S}+a_3^{S}+a_4^{S}
 +a_5^{S}-2a_6^{S}+4a_7^{S}-2a_8^{S}-2a_9^{S})$   \\\hline
 $\Lambda_b^0\to\Sigma^0\eta_1$ & $\frac{1}{6}\lambda_u^\prime(a_1^{S}+a_{10}^{S}-2a_{11}^{S}
 +a_{12}^{S}+3a_{15}^{S}-6a_{16}^{S}+3a_{17}^{S}-2a_2^{S}+a_3^{S}
 +a_4^{S}+a_5^{S}-2a_6^{S}-2a_7^{S}+a_8^{S}+a_9^{S})$  \\\hline
 $\Lambda_b^0\to\Sigma^-\pi^+$ & $\frac{1}{\sqrt{6}}\lambda_c^\prime(a_1^{S,c}-2a_2^{S,c}+a_3^{S,p})
 +\frac{1}{\sqrt{6}}\lambda_t^\prime(2 a_1^{S,p}-a_{18}^{S,p}+2 a_{19}^{S,p}-a_2-a_{20}^{S,p}-3 a_{23}^{S,p}+6 a_{24}^{S,p}-3 a_{25}^{S,p}-a_3^{S,p}$\\
\hline

  \hline
\end{tabular}
\end{table*}
%%%%%%%%%%%%%%%%%%%%%%%%%%%%%%%%%%%%%%%%%%%%%%%%%%%%%%%%%%%%%%%%%%%%%%%%%%
\begin{table*}
\caption{Topological amplitudes of $\Xi_b^{-,0}$ decays into $\mathcal{B}_8^AM$ from the $b\to d$ transition.}\label{ta5}
 \scriptsize
\begin{tabular}{|c|c|c|c|}
\hline\hline
 Channel & Amplitude \\\hline
 $\Xi_b^-\to \Sigma^0\pi^-$ & $-\frac{1}{2}\lambda_c (a^{A,c}_1+a^{A,c}_2-a^{A,c}_4-2 a^{A,c}_6) +\frac{1}{2}\lambda_t(a^{A,p}_1+2 a^{A,p}_{10}+a^{A,p}_{11}-a^{A,p}_{12}-a^{A,p}_{13}+a^{A,p}_{18}+a^{A,p}_{19}
 $\\&$+a^{A,p}_2
 -a_{21}^{A,p}+3 a^{A,p}_{23}+3 a^{A,p}_{24}-3 a^{A,p}_{26}-2 a^{A,p}_{29}-2 a^{A,p}_{30}-6 a^{A,p}_{32}-a^{A,p}_4-a^{A,p}_5+a^{A,p}_7+a^{A,p}_8)$\\&$ -\frac{1}{2}\lambda_u(a^{A}_1-a^{A}_{14}+a^{A}_{18}+a^{A}_{19}+a^{A}_2
 -a^{A}_{21}-2 a^{A}_{28}-2 a^{A}_{30})$  \\\hline
 $\Xi_b^-\to\Sigma^-\pi^0$ & $\frac{1}{2}\lambda_c(a_1^{A,c}+a_2^{A,c}-a_4^{A,c}-2a_6^{A,c}) -\frac{1}{2}\lambda_t(a_1^{A,p}+2a_{10}^{A,p}+a_{11}^{A,p}
 -a_{12}^{A,p}-a_{13}^{A,p}+a_{18}^{A,p}+a_{19}^{A,p}+a_2^{A,p}$\\&$
 -a_{21}^{A,p}+3a_{23}^{A,p}+3a_{24}^{A,p}-3a_{26}^{A,p}-2a_{29}^{A,p}-2a_{30}^{A,p}-6a_{32}^{A,p}-a_4^{A,p}-a_5^{A,p}+a_7^{A,p}+a_8^{A,p}) $\\&$+\frac{1}{2}\lambda_u(a_{13}^{A}+a_{18}^{A}+a_{19}^{A}-a_{21}^{A}+2a_{29}^{A}
 -2a_{30}^{A})$   \\\hline
 $\Xi_b^-\to\Sigma^-\eta_8$ & $\frac{1}{2\sqrt{3}}\lambda_c(-a_1^{A,c}+a_2^{A,c}+2a_3^{A,c}
 +a_4^{A,c}+2a_6^{A,c})-\frac{1}{2\sqrt{3}}\lambda_t(a_1^{A,p}-2a_{10}^{A,p}
 -a_{11}^{A,p}+a_{12}^{A,p}+a_{13}^{A,p}$\\
\hline
  \hline
\end{tabular}
\end{table*}
%%%%%%%%%%%%%%%%%%%%%%%%%%%%%%%%%%%%%%%%%%%%%%%%%%%%%%%%%%%%%%%%%%%%
\begin{table*}
\caption{Topological amplitudes of $\Lambda_b^0$ decays into $\mathcal{B}_8^AM$ from the $b\to d$ transition. }\label{ta6}
 \scriptsize
\begin{tabular}{|c|c|c|c|}
\hline\hline
 Channel & Amplitude \\\hline
  $\Lambda_b^0\to\Sigma^0K^0$ & $\frac{1}{2}\lambda_c(-a_2^{A,c}-a_3^{A,c})+\frac{1}{2}\lambda_t(a_1^{A,p}+a_{19}^{A,p}+a_{20}^{A,p}+3a_{24}^{A,p}+3a_{25}^{A,p}-a_3^{A,p}-a_5^{A,p}-a_6^{A,p}+a_8^{A,p}+a_9^{A,p})$\\&$
 +\frac{1}{2}\lambda_u(-a_{19}^{A}-a_{20}^{A}+a_5^{A}+a_6^{A}
 +a_8^{A}+a_9^{A})$   \\\hline\
 $\Lambda_b^0\to\Sigma^-K^+$ & $\frac{1}{\sqrt{2}}\lambda_c(a_2^{A,c}+a_3^{A,c})-\frac{1}{\sqrt{2}}\lambda_t(a_1^{A,p}+a_{19}^{A,p}+a_{20}^{A,p}+3a_{24}^{A,p}+3a_{25}^{A,p}-a_3^{A,p}-a_5^{A,p}-a_6^{A,p}+a_8^{A,p}$\\&
 $+a_9^{A,p})+\frac{1}{\sqrt{2}}\lambda_u
 (a_{11}^{A}+a_{12}^{A}+a_{19}^{A}+a_{20}^{A})$   \\\hline
 $\Lambda_b^0\to p\pi^-$ & $\frac{1}{\sqrt{2}}\lambda_c(-a_1^{A,c}+a_3^{A,c}+a_4^{A,c}+2a_6^{A,c})+\frac{1}{\sqrt{2}}\lambda_t(2a_{10}^{A,p}+a_{11}^{A,p}-a_{12}^{A,p}-a_{13}^{A,p}+a_{18}^{A,p}+a_2^{A,p}$\\&
 $-a_{20}^{A,p}-a_{21}^{A,p}+3a_{23}^{A,p}-3a_{25}^{A,p}-3a_{26}^{A,p}-2a_{29}^{A,p}+a_3^{A,p}-2a_{30}^{A,p}-6a_{32}^{A,p}-a_4^{A,p}+a_6^{A,p}+a_7^{A,p}-a_9^{A,p})$\\&$
 -\frac{1}{\sqrt{2}}\lambda_u(a_1^{A}-a_{14}^{A}+a_{18}^{A}-a_{20}^{A}-a_{21}^{A}
 -2a_{28}^{A}-a_3^{A}-2a_{30}^{A}+a_4^{A}+a_5^A+a_7^{A}+a_8^{A})$   \\
\hline
 \end{tabular}
\end{table*}
\begin{table*}
\caption{Topological amplitudes of $\Xi_b^{-,0}$ decays into $\mathcal{B}_8^AM$ from the $b\to s$ transition.}\label{ta7}
 \scriptsize
\begin{tabular}{|c|c|c|c|}
\hline\hline
 Channel & Amplitude \\\hline
 $\Xi_b^-\to\Sigma^0K^-$ & $
 \frac{1}{2}\lambda_c^\prime(-a_1^{A,c}+a_3^{A,c}+a_4^{A,c}+2a_6^{A,c})
 +\frac{1}{2}\lambda_t^\prime(2a_{10}^{A,p}+a_{11}^{A,p}-a_{12}^{A,p}-a_{13}^{A,p}
 +a_{18}^{A,p}+a_2^{A,p}-a_{20}^{A,p}-a_{21}^{A,p}+3a_{23}^{A,p}-3a_{25}^{A,p}-3a_{26}^{A,p}$\\&$
 -2a_{29}^{A,p}
 +a_3^{A,p}-2a_{30}^{A,p}-6a_{32}^{A,p}-a_4^{A,p}+a_6^{A,p}+a_7^{A,p}-a_9^{A,p})
 +\frac{1}{2}\lambda_u^\prime(-a_1^A+a_{14}^A
 -a_{18}^A+a_{20}^A+a_{21}^A+2a_{28}^A+a_3^A+2a_{30}^A)$
 \\\hline
 $\Xi_b^-\to\Sigma^-\overline{K}^0$ & $
 \frac{1}{\sqrt{2}}\lambda_c^\prime(-a_1^{A,c}+a_3^{A,c}+a_4^{A,c}+2a_6^{A,c})
 +\frac{1}{\sqrt{2}}\lambda_t^\prime(2a_{10}^{A,p}+a_{11}^{A,p}-a_{12}^{A,p}
 -a_{13}^{A,p}+a_{18}^{A,p}+a_2^{A,p}-a_{20}^{A,p}-a_{21}^{A,p}+3a_{23}^{A,p}
 -3a_{25}^{A,p}$\\&$
 -3a_{26}^{A,p}
 -2a_{29}^{A,p}+a_3^{A,p}-2a_{30}^{A,p}-6a_{32}^{A,p}-a_4^{A,p}+a_6^{A,p}
 +a_7^{A,p}-a_9^{A,p})
 +\frac{1}{\sqrt{2}}\lambda_u^\prime(-a_{18}^A+a_{20}^A+a_{21}^A+2a_{30}^A)$
 \\\hline
 $\Xi_b^-\to\Xi^0\pi^-$ & $
 \frac{1}{\sqrt{2}}\lambda_c^\prime(a_2^{A,c}+a_3^{A,c})
 -\frac{1}{\sqrt{2}}\lambda_t^\prime(a_1^{A,p}+a_{19}^{A,p}+a_{20}^{A,p}
 +3a_{24}^{A,p}+3a_{25}^{A,p}-a_3^{A,p}-a_5^{A,p}-a_6^{A,p}+a_8^{A,p}+a_9^{A,p})
 $\\&$+\frac{1}{\sqrt{2}}\lambda_u^\prime(a_{19}^A +a_2^A+a_{20}^A+a_3^A)$
 \\\hline
 $\Xi_b^-\to\Xi^-\pi^0$ & $
 \frac{1}{2}\lambda_c^\prime(a_2^{A,c}+a_3^{A,c})
 -\frac{1}{2}\lambda_t^\prime(a_1^{A,p}+a_{19}^{A,p}+a_{20}^{A,p}+3a_{24}^{A,p}
 +3a_{25}^{A,p}-a_3^{A,p}-a_5^{A,p}-a_6^{A,p}+a_8^{A,p}+a_9^{A,p})
 $\\
\hline
   \hline
\end{tabular}
\end{table*}

\begin{table*}
\caption{Topological amplitudes of $\Lambda^0_b$ decays into $\mathcal{B}_8^AM$ from the $b\to s$ transition. }\label{ta8}
 \scriptsize
\begin{tabular}{|c|c|c|c|}
\hline\hline
 Channel & Amplitude \\\hline
 $\Lambda_b^0\to\Sigma^+\pi^-$ & $
 \frac{1}{\sqrt{2}}\lambda_c^\prime(a_1^{A,c}-a_3^{A,c})
 -\frac{1}{\sqrt{2}}\lambda_t^\prime(a_{18}^{A,p}+a_2^{A,p}-a_{20}^{A,p}
 +3a_{23}^{A,p}-3a_{25}^{A,p}+a_3^{A,p}-a_4^{A,p}+a_6^{A,p}+a_7^{A,p}
 -a_9^{A,p})$\\&$
 +\frac{1}{\sqrt{2}}\lambda_u^\prime(a_1^A+a_{18}^A-a_{20}^A-a_3^A+a_4^A+a_5^A)$
 \\\hline
 $\Lambda_b^0\to\Sigma^0\pi^0$ & $
 \frac{1}{2\sqrt{2}}\lambda_c^\prime(2a_1^{A,c}-2a_3^{A,c})
 -\frac{1}{\sqrt{2}}\lambda_t^\prime(a_{18}^{A,p}+a_2^{A,p}-a_{20}^{A,p}
 +3a_{23}^{A,p}-3a_{25}^{A,p}+a_3^{A,p}-a_4^{A,p}+a_6^{A,p}+a_7^{A,p}-a_9^{A,p})$\\&$
 +\frac{1}{2\sqrt{2}}\lambda_u^\prime
 (a_1^A+a_{10}^A
 -a_{12}^A+2a_{18}^A-2a_{20}^A-a_3^A+a_4^A+a_5^A)$
 \\\hline
 $\Lambda_b^0\to\Sigma^0\eta_8$ & $
 \frac{1}{2\sqrt{6}}\lambda_u^\prime(a_1^A+a_{10}^A-a_{12}^A-a_3^A-a_4^A-a_5^A-2a_8^A-2a_9^A)$
 \\\hline
 $\Lambda_b^0\to\Sigma^0\eta_1$ & $
 \frac{1}{2\sqrt{3}}\lambda_u^\prime(a_1^A+a_{10}^A-a_{12}^A+3a_{15}^A-3a_{17}^A-a_3^A-a_4^A-a_5^A+a_8^A+a_9^A)$
 \\\hline
 $\Lambda_b^0\to\Sigma^-\pi^+$ & $
 \frac{1}{\sqrt{2}}\lambda_c^\prime(a_1^{A,c}-a_3^{A,c})
 -\frac{1}{\sqrt{2}}\lambda_t^\prime(a_{18}^{A,p}+a_2^{A,p}-a_{20}^{A,p}
 +3a_{23}^{A,p}-3a_{25}^{A,p}+a_3^{A,p}-a_4^{A,p}+a_6^{A,p}+a_7^{A,p}
 -a_9^{A,p})$\\
\hline
    \hline
\end{tabular}
\end{table*}
%%%%%%%%%%%%%%%%%%%%%%%%%%%%%%%%%%%%%%%%%%%%%%%%%%%%%%%%%%%%%%%%%
\begin{table*}
\caption{Decay amplitudes of $\mathcal{B}_{b\overline{3}}\to \mathcal{B}_{8}M$ decays from the $b\to d$ transition, in which the baryon octet is expressed as a $(1,1)$-rank tensor.}\label{ta9}
 \small
\begin{tabular}{|c|c|c|c|}
\hline\hline
 Channel & Amplitude \\\hline
 $\Xi_b^-\to \Sigma^0\pi^-$ & $\frac{1}{\sqrt{2}}\lambda_t(A^\prime_{11}-A^\prime_{12}) +\frac{1}{\sqrt{2}}\lambda_u(A_2+A_3-A_7+A_8)$  \\\hline
 $\Xi_b^-\to \Sigma^-\pi^0$ & $\frac{1}{\sqrt{2}}\lambda_t(-A^\prime_{11}+A^\prime_{12})+\frac{1}{\sqrt{2}}\lambda_u(A_1+A_4+A_7-A_8)$  \\\hline
 $\Xi_b^-\to \Sigma^-\eta_8$ & $-\frac{1}{\sqrt{6}} \lambda_t(A^\prime_{11}+A^\prime_{12})+\frac{1}{\sqrt{6}}\lambda_u(A_1+A_4+A_7+A_8)$  \\\hline
 $\Xi_b^-\to \Sigma^-\eta_1$ & $-\frac{1}{\sqrt{3}}\lambda_t(A^\prime_{11}+A^\prime_{12}+3A^\prime_{13})+\frac{1}{\sqrt{3}}\lambda_u(A_1+3A_{10}+A_4+A_7+A_8)$   \\\hline
 $\Xi_b^-\to \ nK^-$ & $-\lambda_tA^\prime_{11}+\lambda_uA_7$  \\\hline
 $\Xi_b^-\to \Xi^-K^0 $ & $-\lambda_tA^\prime_{12}+\lambda_uA_8$ \\\hline
 $\Xi_b^-\to \Lambda^0\pi^-$ & $-\frac{1}{\sqrt{6}}\lambda_t(A^\prime_{11}+A^\prime_{12})+\frac{1}{\sqrt{6}}\lambda_u(A_2+A_3+A_7+A_8)$  \\
\hline
  \hline
\end{tabular}
\end{table*}
%%%%%%%%%%%%%%%%%%%%%%%%%%%%%%%%%%%%%%%%%%%%%%%%%%%%%%%%%%%%%%%%%%%%%%%%
\begin{table*}
\caption{Decay amplitudes of $\mathcal{B}_{b\overline{3}}\to \mathcal{B}_{8}M$ decays from the $b\to s$ transition, in which the baryon octet is expressed as a $(1,1)$-rank tensor.}\label{ta10}
 \small
\begin{tabular}{|c|c|}
\hline\hline
 Channel & Amplitude \\\hline
$\Xi_b^-\to\Sigma^0K^-$ & $-\frac{1}{\sqrt{2}}\lambda_t^\prime A_{12}^\prime+\frac{1}{\sqrt{2}}\lambda_u^\prime (A_2+A_3+A_8)$   \\\hline
$\Xi_b^-\to\Sigma^-\overline{K}^0$ & $-\lambda_t^\prime A^\prime_{12}+\lambda_u^\prime A_8$   \\\hline
$\Xi_b^-\to\Xi^0\pi^-$ & $-\lambda_t^\prime A_{11}+\lambda_u^\prime A_7$  \\\hline
$\Xi_b^-\to\Xi^-\pi^0$ & $-\frac{1}{\sqrt{2}}\lambda_t^\prime A^\prime_{11}+\frac{1}{\sqrt{2}}\lambda_u^\prime(A_1+A_4+A_7)$   \\\hline
$\Xi_b^-\to\Xi^-\eta_8$ & $-\frac{1}{\sqrt{6}}\lambda_t^\prime(A^\prime_{11}-2A^\prime_{12})+\frac{1}{\sqrt{6}}\lambda_u^\prime(A_1+A_4+A_7-2A_8)$  \\\hline
$\Xi_b^-\to\Xi^-\eta_1$ & $-\frac{1}{\sqrt{3}}\lambda_t^\prime(A^\prime_{11}+A^\prime_{12}+3A^\prime_{13})+\frac{1}{\sqrt{3}}\lambda_u^\prime(A_1+3A_{10}+A_4+A_7+A_8)$   \\\hline
$\Xi_b^-\to\Lambda^0K^-$ & $\frac{1}{\sqrt{6}}\lambda_t^\prime(2A^\prime_{11}-A^\prime_{12})+\frac{1}{\sqrt{6}}\lambda_u^\prime(A_2+A_3-2A_7+A_8)$   \\
\hline
\hline
\end{tabular}
\end{table*}

The topological amplitudes of $\mathcal{B}_{b\overline 3}\to \mathcal{B}_8M$ decays are presented in Tables~\ref{ta1}$\sim$\ref{ta8}.
We use the superscript $P$ to distinguish the penguin-induced diagrams from the tree-induced diagrams.
The decay amplitudes constructed by $(\mathcal{B}_8)^i_j$ are listed in Tables~\ref{ta9} and \ref{ta10}.
Note that the tree-induced amplitudes $A_{15}$ $\sim$ $A_{18}$ and penguin-induced amplitudes $A_{1}^P$ and $A_{10}^P$ do not contribute to the bottom baryon decays.
One can verify that Tables~\ref{ta9} and \ref{ta10} are consistent with Tables~\ref{ta1} to \ref{ta8} using the decomposition in Eq.~\eqref{sol}.
For example, the topological amplitudes contributing to the $\Lambda^0_b\to p\pi^-$ decay are
\begin{align}
  \mathcal{A}(\Lambda^0_b\to p \pi^-) &= \lambda_u A_2-\lambda_t A^\prime_{12} \nonumber\\&=\lambda_u (A_2+A_{12})+\lambda_cA^c_{1}-\lambda_t(A^P_{4}+A^P_{6}+A^P_{12}+3A^P_{15})\nonumber\\ & = \frac{1}{\sqrt{2}}\lambda_u(-a^A_1+a^A_{3}-a^A_{4}-a^A_{5}
  -a^A_7-a^A_8+a^A_{14}-a^A_{18}+a^A_{20}+a^A_{21}+2a^A_{28}+2a^A_{30})\nonumber\\&~~
  -\frac{1}{\sqrt{6}}\lambda_u(a^S_1-2a^S_{2}+a^S_{3}-a^S_{4}-a^S_{5}
 +2a^S_{6}-a^S_7-a^S_{8}+2a^S_{9}+3a^S_{14}+a^S_{18}-2a^S_{19}
 +a^S_{20}+3a^S_{21})\nonumber\\&~~+\frac{1}{\sqrt{2}}(-a^{c,A}_{1}+a^{c,A}_{3}+a^{c,A}_{4}
 +2a^{c,A}_{6})-\frac{1}{\sqrt{6}}(
 a^{c,S}_{1}-2a^{c,S}_{2}
 +a^{c,S}_{3}+3a^{c,S}_{4})\nonumber\\&~~-\frac{1}{\sqrt{2}}\lambda_t(-a_2^{A,p}
 -a_3^{A,p}+a_4^{A,p}-a_6^{A,p}
 -a_7^{A,p}+a_9^{A,p}-2a_{10}^{A,p}
 -a_{11}^{A,p}+a_{12}^{A,p}+a_{13}^{A,p}
 -a_{18}^{A,p}+a_{20}^{A,p}\nonumber\\&~~+a_{21}^{A,p}
 -3a_{23}^{A,p}+3a_{25}^{A,p}+3a_{26}^{A,p}
 +2a_{29}^{A,p}+2a_{30}^{A,p}+6a_{32}^{A,p})
 -\frac{1}{\sqrt{6}}\lambda_t(-2a_1^{S,p}+a_2^{S,p}+a_3^{S,p}\nonumber\\&~~
 -a_4^{S,p}+2a_5^{S,p}-a_6^{S,p}+a_7^{S,p}-2a_8^{S,p}+a_9^{S,p}
 -3a_{11}^{S,p}+3a_{12}^{S,p}+3a_{13}^{S,p}+a_{18}^{S,p}-2a_{19}^{S,p}+a_{20}^{S,p}
 \nonumber\\&~~+3a_{21}^{S,p}+3a_{23}^{S,p}-6a_{24}^{S,p}+3a_{25}^{S,p}+9a_{26}^{S,p}).
\end{align}
Moreover, one can verify the isospin relations derived in Ref.~\cite{Fu:2025wwx} and the $U$-spin or $SU(3)_F$ relations provided in Refs.~\cite{He:2015fwa,He:2015fsa,Roy:2019cky,Roy:2020nyx} using Tables~\ref{ta9} and \ref{ta10}.

The linear correlation of decay amplitudes contributing to the $\mathcal{B}_{b\overline 3}\to \mathcal{B}_8M$ modes in the Standard Model is beyond the model-independent analysis in Appendix.~\ref{su3}.
The nonzero coefficients induced by tree operators $\mathcal{O}^{(u)}_{1,2}$ in the $SU(3)$ irreducible representations are
\begin{align}\label{ckm3}
& H(\overline 6)^{(u)31} = -V_{ub}V^*_{ud}, \qquad H(\overline 6)^{(u)12} = V_{ub}V^*_{us},\qquad H(15)_{12}^{(u)1} = 3V_{ub}V^*_{ud}, \qquad H(15)_{22}^{(u)2} = -2V_{ub}V^*_{ud},\nonumber\\& H(15)_{32}^{(u)3} = -V_{ub}V^*_{ud},\qquad H(15)_{13}^{(u)1} = 3V_{ub}V^*_{us},\qquad H(15)_{33}^{(u)3} = -2V_{ub}V^*_{us}, \qquad H(15)_{32}^{(u)2} = -V_{ub}V^*_{us}\nonumber\\
& H^{(u)}(3_t)_2 = V_{ub}V^*_{ud}, \qquad H^{(u)}(3_t)_3 =  V_{ub}V^*_{us}.
\end{align}
The nonzero coefficients induced by tree operators $\mathcal{O}^{(c)}_{1,2}$ in the $SU(3)$ irreducible representations are
\begin{align}
H(3)^{(c)}_2 = V_{cb}V^*_{cd}, \qquad H(3)^{(c)}_3 = V_{cb}V^*_{cs}.
\end{align}
The nonzero coefficients induced by penguin operators in the $SU(3)$ irreducible representations are
\begin{align}\label{ckm4}
 H(3_t)^{(p)}_2=-V_{tb}V_{td}^*, \qquad H(3_p)^{(p)}_2=-3V_{tb}V_{td}^*,\qquad H(3_t)^{(p)}_3=-V_{tb}V_{ts}^*, \qquad H(3_p)^{(p)}_3=-3V_{tb}V_{ts}^*.
\end{align}
According to Eqs.~\eqref{ckm3}$\sim$\eqref{ckm4}, there are no penguin-induced amplitudes in the $15$ and $\overline 6$ irreducible representations.
Furthermore, only the second component of $3$-dimensional presentation in the $\Delta S =0$ transition is non-zero.
Therefore, the amplitudes induced by the $3$-dimensional representations always appear simultaneously and can be absorbed into four amplitudes,
\begin{align}\label{c5}
b^\prime_{11} & = -\frac{\lambda_u}{\lambda_t}b_{15}-\frac{\lambda_c}{\lambda_t}b^c_{4}+3b^P_{11} + b^P_{15},  \qquad   b^\prime_{12} =-\frac{\lambda_u}{\lambda_t}b_{16}-\frac{\lambda_c}{\lambda_t}b^c_{2}+3b^P_{12} + b^P_{12}, \nonumber\\  b^\prime_{13} & = -\frac{\lambda_u}{\lambda_t}b_{17}-\frac{\lambda_c}{\lambda_t}b^c_{3}+3b^P_{13} + b^P_{17}, \qquad   b^\prime_{14} = -\frac{\lambda_u}{\lambda_t}b_{18}-\frac{\lambda_c}{\lambda_t}b^c_{1}+3 b^P_{14} + b^P_{18},
\end{align}
where $\lambda_u = V_{ub}V^*_{ud}$, $\lambda_c = V_{cb}V^*_{cd}$, $\lambda_t = V_{tb}V^*_{td}$.
According to Eqs.~\eqref{c5} and \eqref{sol2}, the tree-induced amplitudes $A_{11} \sim A_{14}$, $A^c_{1} \sim A^c_{4}$, and all the penguin-induced amplitudes can be absorbed into four amplitudes in the $SU(3)_F$ limit,
\begin{align}\label{c11}
 & A^{\prime}_{11}  = -\frac{\lambda_u}{\lambda_t}A_{11}-\frac{\lambda_c}{\lambda_t}A^c_{3}+A^P_{3}+A^P_{5}+A^P_{11}+3A^P_{17},\qquad A^{\prime}_{12}  = -\frac{\lambda_u}{\lambda_t}A_{12}-\frac{\lambda_c}{\lambda_t}A^c_{1}+A^P_{4}+A^P_{6}+A^P_{12}+3A^P_{15},\nonumber\\
& A^{\prime}_{13}  = -\frac{\lambda_u}{\lambda_t}A_{13}-\frac{\lambda_c}{\lambda_t}A^c_{2}+A^P_{2}+A^P_{9}+A^P_{13}+3A^P_{16},\qquad A^{\prime}_{14}  = -\frac{\lambda_u}{\lambda_t}A_{14}-\frac{\lambda_c}{\lambda_t}A^c_{4}+A^P_{7}+A^P_{8}+A^P_{14}+3A^P_{18}.
\end{align}
For the $\Delta S =-1$ transition, $\lambda_{u,c,t}$ in Eqs.~\eqref{c5} and \eqref{c11} are replaced by  $\lambda^\prime_{u,c,t}$ and $\lambda^\prime_u = V_{ub}V^*_{us}$, $\lambda^\prime_c = V_{cb}V^*_{cs}$, $\lambda^\prime_t = V_{tb}V^*_{ts}$.
According to Eq.~\eqref{c11}, all the penguin induced amplitudes are determined once the tree induced amplitudes with quark loops are known.
There is no degree of freedom of penguin induced diagrams in the SM.
Combining Eqs.~\eqref{c9} and \eqref{c5}, there are $13$ independent amplitudes contributing to the $\mathcal{B}_{b\overline 3}\to \mathcal{B}_8M$ decays in the $SU(3)_F$ limit.

\section{Phenomenological analysis}\label{pa}

First, we compare the topological amplitudes in bottom and charmed baryon decays.
For charmed baryon decays, the contributions from the $d$- and $s$-quark loop diagrams cancel each other out because $V_{cd}^*V_{ud}\sim -V_{cs}^*V_{us}$.
The penguin-induced amplitudes are much smaller than the tree-induced amplitudes under the $SU(3)_F$ limit, as $|V_{cb}^*V_{ub}/V_{cd/s}^*V_{ud/s}|\times (\alpha_s/\pi)\sim 10^{-4}$.
Note that all the penguin-induced diagrams and tree-induced diagrams with quark-loops are included in $A^\prime_{11}\sim A^{\prime}_{14}$.
Consequently, $A_{11}^{\prime}\sim A_{14}^{\prime}$ are negligible in the branching fractions of charmed baryon decays.
For bottom baryon decays, the $b\to d$ transition is dominated by tree-induced amplitudes, as $|V_{tb}V_{td}^*/V_{ub}V_{ud}^*|\times (\alpha_s/\pi)\sim 0.1$.
The $b\to s$ transition is dominated by the penguin-induced and quark-loop diagrams, since $|V_{tb}V_{ts}^*/V_{ub}V_{us}^*|\times (\alpha_s/\pi)\sim 2$.
The amplitudes $A_{11}^{\prime}\sim A_{14}^{\prime}$ are essential to the branching fractions of bottom baryon decays.
The CP asymmetries in the bottom baryon decays arise from the interference between tree amplitudes $A_1\sim A_{10}$ and $A^\prime_{11}\sim A^{\prime}_{14}$.
The CP asymmetries of charmless bottom baryon decays could reach $\mathcal{O}(0.1)$, whereas the CP asymmetries of charmed bottom baryon decays are estimated to be $\mathcal{O}(10^{-4})$.
Because of the different CKM matrix elements, the $U$-spin relations between two $U$-spin conjugate modes differ for charmed and bottom baryon decays \cite{Zhang:2025jnw}.

Owing to the limited experimental data, we cannot determine decay amplitudes by global fitting.
However, some dynamical information can still be extracted from the available data.
The CP asymmetries in the $\Lambda^0_b\to p\pi^-$ and $\Lambda^0_b\to pK^-$ decays have been measured by the LHCb Collaboration as \cite{LHCb:2024iis}
\begin{align}
 A_{CP}(\Lambda^0_b\to p\pi^-) = (0.2\pm0.8\pm0.4)\%,\qquad\qquad A_{CP}(\Lambda^0_b\to pK^-) = (-1.1\pm0.7\pm0.4)\%.
\end{align}
The branching fractions for these decays are given by \cite{ParticleDataGroup:2024cfk}
\begin{align}
 \mathcal{B}r(\Lambda^0_b\to p\pi^-) = (4.6\pm0.8)\times 10^{-6},\qquad\qquad \mathcal{B}r(\Lambda^0_b\to pK^-) = (5.5\pm1.0)\times 10^{-6}.
\end{align}
The branching fractions of the $\Lambda^0_b\to p\pi^-$ and $\Lambda^0_b\to pK^-$ decays are of the same order, indicting that penguin amplitudes are significant.
The small CP asymmetries in these two decay modes, particularly in $\Lambda^0_b\to pK^-$, might result from small relative phases between the tree and penguin amplitudes.
According to the PQCD calculations \cite{Han:2024kgz}, the color-favored emitted diagram $T$, denoted as $a^A_{28}$ in this paper, is one order larger than the other diagrams.
Eq.~\eqref{sol} shows that $a^A_{28}$ contributes only to the amplitude $A_2$.
Consequently, $A_2$ is expected to be larger than other amplitudes.
The small CP violation in the $\Lambda^0_b\to pK^-$ mode might be explained by small relative strong phases $\delta_{A_{12}^{\prime}}-\delta_{A_{2}}$ and $\delta_{A_{14}^{\prime}}-\delta_{A_{2}}$.
Hence, we suggest measuring CP asymmetries in the $\Xi^0_b\to pK^-$ and $\Xi^-_b\to \Lambda^0 K^-$ decays on LHCb.
In these processes, the tree and penguin amplitudes may be of comparable magnitude after accounting for the CKM matrix elements, and the small strong phases $\delta_{A_{12}^{\prime}}-\delta_{A_{2}}$ and $\delta_{A_{14}^{\prime}}-\delta_{A_{2}}$ are avoided.

In the literature, the K\"orner-Pati-Woo theorem \cite{Pati:1970fg,Korner:1970xq} plays an important role in analyzing heavy baryon weak decays.
It states that the two quarks produced by weak operators must be antisymmetric in flavor if they enter the same low-lying baryon.
In diagrams $a_1^{S,A}\sim a_{12}^{S,A}$, $a_{15}^{S,A}\sim a_{17}^{S,A}$, the two quarks emitted from the weak vertex enter the final-state or resonance-state baryon.
They should be antisymmetric in flavor if the K\"orner-Pati-Woo theorem holds.
If two of the diagrams $a_1^{S,A}\sim a_{12}^{S,A}$, $a_{15}^{S,A}\sim a_{17}^{S,A}$ are connected with each other by interchanging two emitted quarks, these two diagrams are opposite.
Then, we have
\begin{align}
  & a^{S,A}_1 = -a^{S,A}_2,\qquad   a^{S,A}_5 = -a^{S,A}_6, \quad a^{S,A}_7 = -a^{S,A}_{10}, \\ &  a^{S,A}_8 = -a^{S,A}_{11}, \qquad a^{S,A}_9  = -a^{S,A}_{12}, \qquad a^{S,A}_{15}  = -a^{S,A}_{16}.
\end{align}
Furthermore, the antisymmetry of the two emitted quarks conflicts with the requirement that these two quarks are symmetric in diagrams $a^{S}_{3},a^{S}_{4},a^{S}_{17}$, which results in
\begin{align}
  a^{S}_{3} =a^{S}_{4}= a^{S}_{17} = 0.
\end{align}
Applying the above equations to Eq.~\eqref{sol}, we obtain the following relations:
\begin{align}
  A_1=-A_3,\qquad A_5=-A_7,\qquad A_6=-A_8,\qquad A_9=-A_{10}.
\end{align}
The decay amplitudes of the $\Lambda_b^0\to\Sigma^0K^0$ and $\Lambda_b^0\to\Sigma^-K^+$ modes are
\begin{align}
  \mathcal{A}(\Lambda_b^0\to\Sigma^0K^0)
  =\frac{1}{\sqrt{2}}\lambda_tA^\prime_{11}+\frac{1}{\sqrt{2}}\lambda_uA_3,\qquad
  \mathcal{A}(\Lambda_b^0\to\Sigma^-K^+)=-\lambda_tA^\prime_{11}+\lambda_uA_1.
\end{align}
If the K\"orner-Pati-Woo theorem is valid, the relation $A_1=-A_3$ leads to
\begin{align}\label{q1}
  \mathcal{B}r(\Lambda_b^0\to\Sigma^-K^+)= 4\,\mathcal{B}r(\Lambda_b^0\to\Sigma^0K^0_S)
\end{align}
under the isospin symmetry.
Isospin breaking is naively predicted to be $\mathcal{O}(1\%)$.
If Eq.~\eqref{q1} is violated, it indicates that the K\"orner-Pati-Woo theorem is not valid in topological diagrams.

\section{Summary}\label{summary}
In summary, we studied the topological amplitudes of $\mathcal{B}_{b\overline 3}\to \mathcal{B}_8 M$ decays in the $SU(3)_F$ limit.
The linear relations between topologies and $SU(3)$ irreducible amplitudes are derived through tensor contraction and $SU(3)$ decomposition.
Thirteen independent amplitudes contribute to $\mathcal{B}_{b\overline{3}}\to \mathcal{B}_8 M$ decays in the Standard Model, of which four are crucial to the CP asymmetries.
The small CP asymmetries in the $\Lambda^0_b\to p\pi^-$ and $\Lambda^0_b\to pK^-$ modes might indicate small strong phases between amplitudes $A_2$ and $A_{12,14}^\prime$.
To avoid the small strong phases, we suggest measuring CP asymmetries in the $\Xi^0_b\to pK^-$ and $\Xi^-_b\to \Lambda^0 K^-$ decays.
Moreover, the K\"orner-Pati-Woo theorem can be tested by measuring the branching fractions of the $\Lambda_b^0\to\Sigma^0K^0_S$ and $\Lambda_b^0\to\Sigma^-K^+$ decays under isospin symmetry.

\begin{acknowledgements}

We are grateful to Fu-Sheng Yu for useful discussions.
This work was supported in part by the National Natural Science Foundation of China under Grants No. 12105099.

\end{acknowledgements}

\begin{appendix}
%%%%%%%%%%%%%%%%%%%%%%%%%%%%%%%
\section{Topological amplitudes of bottom baryon decays}\label{su3}

The bottom baryon anti-triplet is
\begin{eqnarray}
 |\mathcal{B}_{b\overline 3}\rangle=  \left( \begin{array}{ccc}
   0   & \Lambda_b^0  & \Xi_b^0 \\
    -\Lambda_b^0 &   0   & \Xi_b^- \\
    -\Xi_b^0 & -\Xi_b^- & 0 \\
  \end{array}\right).
\end{eqnarray}
The light pseudoscalar meson nonet is
\begin{eqnarray}
 |M\rangle=  \left( \begin{array}{ccc}
   \frac{1}{\sqrt 2} \pi^0+  \frac{1}{\sqrt 6} \eta_8    & \pi^+  & K^+ \\
    \pi^- &   - \frac{1}{\sqrt 2} \pi^0+ \frac{1}{\sqrt 6} \eta_8   & K^0 \\
    K^- & \overline K^0 & -\sqrt{2/3}\eta_8 \\
  \end{array}\right) +  \frac{1}{\sqrt 3} \left( \begin{array}{ccc}
   \eta_1    & 0  & 0 \\
    0 &  \eta_1   & 0 \\
   0 & 0 & \eta_1 \\
  \end{array}\right).
\end{eqnarray}
The light baryon octets $\mathcal{B}_8^S$ and $\mathcal{B}_8^A$ can be written as tensors with three covariant indices.
$\mathcal{B}_8^S$ is given by
\begin{align}\label{BS}
  \Sigma^+ & = \frac{1}{\sqrt{6}}(2\mathcal{B}_8^{113}-\mathcal{B}_8^{311}-\mathcal{B}_8^{131}),
  \qquad   p =\frac{1}{\sqrt{6}}(-2\mathcal{B}_8^{112}+\mathcal{B}_8^{211}+\mathcal{B}_8^{121}), \qquad   \Sigma^- = \frac{1}{\sqrt{6}}(-2\mathcal{B}_8^{223}+\mathcal{B}_8^{322}+\mathcal{B}_8^{232}), \nonumber\\
    n &=\frac{1}{\sqrt{6}}(2\mathcal{B}_8^{221}-\mathcal{B}_8^{122}-\mathcal{B}_8^{212}), \qquad\Xi^- = \frac{1}{\sqrt{6}}(2\mathcal{B}_8^{332}-\mathcal{B}_8^{233}-\mathcal{B}_8^{323}), \qquad   \Xi^0 =\frac{1}{\sqrt{6}}(-2\mathcal{B}_8^{331}+\mathcal{B}_8^{133}+\mathcal{B}_8^{313}),\nonumber\\
 \Sigma^0 &=\frac{1}{\sqrt{12}}(\mathcal{B}_8^{321}+\mathcal{B}_8^{231}+\mathcal{B}_8^{312}
 +\mathcal{B}_8^{132}-2\mathcal{B}_8^{123}-2\mathcal{B}_8^{213}), \qquad \Lambda^0 = \frac{1}{2}(\mathcal{B}_8^{132}-\mathcal{B}_8^{231}+\mathcal{B}_8^{312}-\mathcal{B}_8^{321}
 ),
\end{align}
and $\mathcal{B}_8^A$ is given by
\begin{align}\label{BA}
  \Sigma^+ & = \frac{1}{\sqrt{2}}(\mathcal{B}_8^{311}-\mathcal{B}_8^{131}),\qquad   p =\frac{1}{\sqrt{2}}(\mathcal{B}_8^{121}-\mathcal{B}_8^{211}), \qquad \Sigma^- = \frac{1}{\sqrt{2}}(\mathcal{B}_8^{232}-\mathcal{B}_8^{322}),  \nonumber\\
  n& =\frac{1}{\sqrt{2}}(\mathcal{B}_8^{122}-\mathcal{B}_8^{212}), \qquad  \Xi^- = \frac{1}{\sqrt{2}}(\mathcal{B}_8^{233}-\mathcal{B}_8^{323}),\qquad   \Xi^0 =\frac{1}{\sqrt{2}}(\mathcal{B}_8^{313}-\mathcal{B}_8^{133}),  \nonumber\\
\Sigma^0 &=\frac{1}{2}(\mathcal{B}_8^{132}-\mathcal{B}_8^{312}+\mathcal{B}_8^{231}
-\mathcal{B}_8^{321}),   \qquad   \Lambda^0  = \frac{1}{\sqrt{12}}(2\mathcal{B}_8^{213}-2\mathcal{B}_8^{123}+\mathcal{B}_8^{231}
-\mathcal{B}_8^{321}+\mathcal{B}_8^{312}-\mathcal{B}_8^{132}).
\end{align}
The decay amplitude of the $\mathcal{B}_{b\overline 3}\to \mathcal{B}_8M$ decay can be constructed via Eq.~\eqref{amp}.
The number of topologies can be counted by permutation.
For the contributions from operators $\mathcal{O}^{(u,p)}$,
there are five covariant/contravariant indices in $\langle \mathcal{B}_{8}^{ijk} M^l_m |\mathcal{O}_{np}^{(u,p)q}|[\mathcal{B}_{c\overline 3}]_{rs}\rangle$.
The number of full contractions is $N=A^5_5=120$.
Considering that the two flavor indices in the antitriplet are antisymmetric, the number of terms is $N_S+N_A=N/2=60$.
As the symmetric light quarks in $\mathcal{B}^S_8$ conflict with the antisymmetric light quarks in $\mathcal{B}_{b\overline 3}$, the number of topological diagrams for bottom antitriplet baryon decays into $\mathcal{B}^S_8$ is less than that for decays into $\mathcal{B}^A_8$.
The difference between them, $N_A-N_S$, is computed to be $A_{3}^3 = 6$.
Then $N_S$ and $N_A$ are solved to be $N_S=27$, $N_A=33$.
For the contributions from $\mathcal{O}^{(c)}$, there are four covariant/contravariant indices in $\langle \mathcal{B}_{8}^{ijk} M^l_m |\mathcal{O}^{(c)}_{n}|[\mathcal{B}_{b\overline 3}]_{op}\rangle$.
As $N^c_S+N^c_A=A^4_4/2=12$ and $N^c_A-N^c_S=A^2_2=2$, we have $N^c_S=5$ and $N^c_A=7$.

\begin{figure}
  \centering
  \includegraphics[width=14cm]{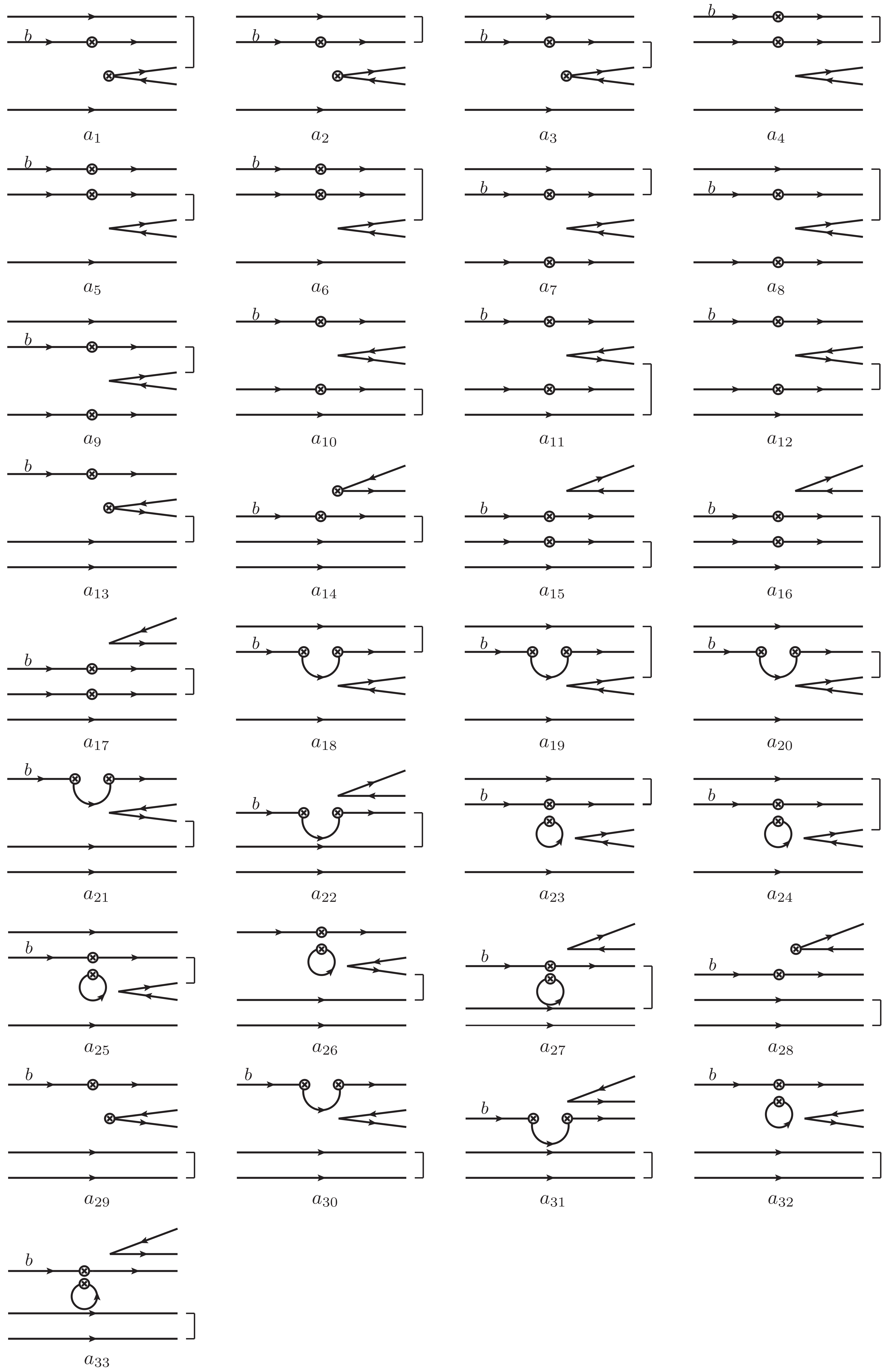}
  \caption{Topological diagrams of bottom baryon $\mathcal{B}_{b\overline 3}$ decays into a light baryon octet $\mathcal{B}^A_8$ and a light meson $M$, in which $"]"$ indicates two antisymmetric light quarks in baryons.}\label{top7}
\end{figure}

The topological amplitude of the $\mathcal{B}_{b\overline 3}\to \mathcal{B}_8^SM$ decay is constructed as
\begin{align}\label{amp1}
 \mathcal{A}^S(\mathcal{B}_{b\overline 3}\to \mathcal{B}_8^SM) = & a^S_1(\mathcal{B}_{b\overline 3})_{ij} H^m_{kl}M^i_m ( \mathcal{B}_8^S)^{jkl} + a^S_2(\mathcal{B}_{b\overline 3})_{ij}H^m_{kl}M^i_m ( \mathcal{B}_8^S)^{jlk}+ a^S_3(\mathcal{B}_{b\overline 3})_{ij}H^m_{kl}M^i_m ( \mathcal{B}_8^S)^{klj} \nonumber\\
   & + a^S_4(\mathcal{B}_{b\overline 3})_{ij} H^i_{kl}M^j_m ( \mathcal{B}_8^S)^{klm}  + a^S_5(\mathcal{B}_{b\overline 3})_{ij} H^i_{kl}M^j_m ( \mathcal{B}_8^S)^{kml}  + a^S_6(\mathcal{B}_{b\overline 3})_{ij} H^i_{kl}M^j_m ( \mathcal{B}_8^S)^{lmk} \nonumber\\
   &+ a^S_7(\mathcal{B}_{b\overline 3})_{ij} H^i_{kl}M^k_m ( \mathcal{B}_8^S)^{jlm}+ a^S_8(\mathcal{B}_{b\overline 3})_{ij} H^i_{kl}M^k_m ( \mathcal{B}_8^S)^{jml} + a^S_9(\mathcal{B}_{b\overline 3})_{ij} H^i_{kl}M^k_m ( \mathcal{B}_8^S)^{lmj} \nonumber\\
 &+ a^S_{10}(\mathcal{B}_{b\overline 3})_{ij} H^i_{kl}M^l_m ( \mathcal{B}_8^S)^{jkm}  + a^S_{11}(\mathcal{B}_{b\overline 3})_{ij} H^i_{kl}M^l_m ( \mathcal{B}_8^S)^{jmk} + a^S_{12}(\mathcal{B}_{b\overline 3})_{ij} H^i_{kl}M^l_m ( \mathcal{B}_8^S)^{kmj} \nonumber\\
&+ a^S_{13}(\mathcal{B}_{b\overline 3})_{ij} H^m_{kl}M^l_m (\mathcal{B}_8^S)^{ikj}+ a^S_{14}(\mathcal{B}_{b\overline 3})_{ij} H^m_{kl}M^k_m ( \mathcal{B}_8^S)^{ilj} + a^S_{15}(\mathcal{B}_{b\overline 3})_{ij} H^i_{kl}M^m_m ( \mathcal{B}_8^S)^{jkl} \nonumber\\
 &+ a^S_{16}(\mathcal{B}_{b\overline 3})_{ij} H^i_{kl}M^m_m (\mathcal{B}_8^S)^{jlk} + a^S_{17}(\mathcal{B}_{b\overline 3})_{ij} H^i_{kl}M^m_m ( \mathcal{B}_8^S)^{klj}  + a^S_{18}(\mathcal{B}_{b\overline 3})_{ij} H^l_{kl}M^i_m ( \mathcal{B}_8^S)^{jkm}\nonumber\\
 & + a^S_{19}(\mathcal{B}_{b\overline 3})_{ij} H^l_{kl}M^i_m (\mathcal{B}_8^S)^{jmk} + a^S_{20}(\mathcal{B}_{b\overline 3})_{ij} H^l_{kl}M^i_m (\mathcal{B}_8^S)^{kmj} + a^S_{21}(\mathcal{B}_{b\overline 3})_{ij} H^l_{kl}M^k_m ( \mathcal{B}_8^S)^{imj}\nonumber\\
& + a^S_{22}(\mathcal{B}_{b\overline 3})_{ij} H^l_{kl}M^m_m (\mathcal{B}_8^S)^{ikj} + a^S_{23}(\mathcal{B}_{b\overline 3})_{ij} H^l_{lk}M^i_m ( \mathcal{B}_8^S)^{jkm}+ a^S_{24}(\mathcal{B}_{b\overline 3})_{ij} H^l_{lk}M^i_m ( \mathcal{B}_8^S)^{jmk}\nonumber\\
& + a^S_{25}(\mathcal{B}_{b\overline 3})_{ij} H^l_{lk}M^i_m (\mathcal{B}_8^S)^{kmj} + a^S_{26}(\mathcal{B}_{b\overline 3})_{ij} H^l_{lk}M^k_m ( \mathcal{B}_8^S)^{imj}+ a^S_{27}(\mathcal{B}_{b\overline 3})_{ij} H^l_{lk}M^m_m ( \mathcal{B}_8^S)^{ikj}\nonumber\\& +a^{c,S}_{1}(\mathcal{B}_{b\overline 3})_{ij} H^{(c)}_{k}M^i_l ( \mathcal{B}_8^S)^{jkl}+ a^{c,S}_{2}(\mathcal{B}_{b\overline 3})_{ij} H^{(c)}_{k}M^i_l ( \mathcal{B}_8^S)^{jlk}+ a^{c,S}_{3}(\mathcal{B}_{b\overline 3})_{ij} H^{(c)}_{k}M^i_l (\mathcal{B}_8^S)^{klj} \nonumber\\
& + a^{c,S}_{4}(\mathcal{B}_{b\overline 3})_{ij} H^{(c)}_{k}M^k_l ( \mathcal{B}_8^S)^{ilj}+ a^{c,S}_{5}(\mathcal{B}_{c\overline 3})_{ij} H^{(c)}_{k}M^l_l ( \mathcal{B}_8^S)^{ikj}.
\end{align}
The topological amplitude of the $\mathcal{B}_{b\overline 3}\to \mathcal{B}_8^AM$ decay is constructed as
\begin{align}\label{amp2}
 \mathcal{A}^A(\mathcal{B}_{b\overline 3}\to \mathcal{B}_8^AM) = & a^A_1(\mathcal{B}_{b\overline 3})_{ij} H^m_{kl}M^i_m ( \mathcal{B}_8^A)^{jkl} + a^A_2(\mathcal{B}_{b\overline 3})_{ij}H^m_{kl}M^i_m ( \mathcal{B}_8^A)^{jlk}+ a^A_3(\mathcal{B}_{b\overline 3})_{ij}H^m_{kl}M^i_m ( \mathcal{B}_8^A)^{klj} \nonumber\\
   & + a^A_4(\mathcal{B}_{b\overline 3})_{ij} H^i_{kl}M^j_m (\mathcal{B}_8^A)^{klm}  + a^A_5(\mathcal{B}_{b\overline 3})_{ij} H^i_{kl}M^j_m (\mathcal{B}_8^A)^{kml}  + a^A_6(\mathcal{B}_{b\overline 3})_{ij} H^i_{kl}M^j_m ( \mathcal{B}_8^A)^{lmk} \nonumber\\
   &+ a^A_7(\mathcal{B}_{b\overline 3})_{ij} H^i_{kl}M^k_m (\mathcal{B}_8^A)^{jlm}+ a^A_8(\mathcal{B}_{b\overline 3})_{ij} H^i_{kl}M^k_m (\mathcal{B}_8^A)^{jml} + a^A_9(\mathcal{B}_{b\overline 3})_{ij} H^i_{kl}M^k_m ( \mathcal{B}_8^A)^{lmj} \nonumber\\
 &+ a^A_{10}(\mathcal{B}_{b\overline 3})_{ij} H^i_{kl}M^l_m ( \mathcal{B}_8^A)^{jkm}  + a^A_{11}(\mathcal{B}_{b\overline 3})_{ij} H^i_{kl}M^l_m (\mathcal{B}_8^A)^{jmk} + a^A_{12}(\mathcal{B}_{b\overline 3})_{ij} H^i_{kl}M^l_m ( \mathcal{B}_8^A)^{kmj} \nonumber\\
&+ a^A_{13}(\mathcal{B}_{b\overline 3})_{ij} H^m_{kl}M^l_m ( \mathcal{B}_8^A)^{ikj}+ a^A_{14}(\mathcal{B}_{b\overline 3})_{ij} H^m_{kl}M^k_m (\mathcal{B}_8^A)^{ilj} + a^A_{15}(\mathcal{B}_{b\overline 3})_{ij} H^i_{kl}M^m_m ( \mathcal{B}_8^A)^{jkl} \nonumber\\
 &+ a^A_{16}(\mathcal{B}_{b\overline 3})_{ij} H^i_{kl}M^m_m ( \mathcal{B}_8^A)^{jlk} + a^A_{17}(\mathcal{B}_{b\overline 3})_{ij} H^i_{kl}M^m_m (\mathcal{B}_8^A)^{klj}  + a^A_{18}(\mathcal{B}_{b\overline 3})_{ij} H^l_{kl}M^i_m ( \mathcal{B}_8^A)^{jkm}\nonumber\\
 & + a^A_{19}(\mathcal{B}_{b\overline 3})_{ij} H^l_{kl}M^i_m ( \mathcal{B}_8^A)^{jmk} + a^A_{20}(\mathcal{B}_{b\overline 3})_{ij} H^l_{kl}M^i_m ( \mathcal{B}_8^A)^{kmj} + a^A_{21}(\mathcal{B}_{b\overline 3})_{ij} H^l_{kl}M^k_m ( \mathcal{B}_8^A)^{imj}\nonumber\\
& + a^A_{22}(\mathcal{B}_{b\overline 3})_{ij} H^l_{kl}M^m_m ( \mathcal{B}_8^A)^{ikj} + a^A_{23}(\mathcal{B}_{b\overline 3})_{ij} H^l_{lk}M^i_m ( \mathcal{B}_8^A)^{jkm}+ a^A_{24}(\mathcal{B}_{b\overline 3})_{ij} H^l_{lk}M^i_m ( \mathcal{B}_8^A)^{jmk}\nonumber\\
& + a^A_{25}(\mathcal{B}_{b\overline 3})_{ij} H^l_{lk}M^i_m ( \mathcal{B}_8^A)^{kmj} + a^A_{26}(\mathcal{B}_{b\overline 3})_{ij} H^l_{lk}M^k_m ( \mathcal{B}_8^A)^{imj}+ a^A_{27}(\mathcal{B}_{b\overline 3})_{ij} H^l_{lk}M^m_m ( \mathcal{B}_8^A)^{ikj}\nonumber\\
&+ a^A_{28}(\mathcal{B}_{b\overline 3})_{ij} H^m_{kl}M^k_m (\mathcal{B}_8^A)^{ijl} + a^A_{29}(\mathcal{B}_{b\overline 3})_{ij} H^m_{kl}M^l_m (\mathcal{B}_8^A)^{ijk} + a^A_{30}(\mathcal{B}_{b\overline 3})_{ij} H^l_{kl}M^k_m ( \mathcal{B}_8^A)^{ijm} \nonumber\\
&+ a^A_{31}(\mathcal{B}_{b\overline 3})_{ij} H^l_{kl}M^m_m ( \mathcal{B}_8^A)^{ijk}+ a^A_{32}(\mathcal{B}_{b\overline 3})_{ij} H^l_{lk}M^k_m ( \mathcal{B}_8^A)^{ijm}+ a^A_{33}(\mathcal{B}_{b\overline 3})_{ij} H^l_{lk}M^m_m ( \mathcal{B}_8^A)^{ijk}\nonumber\\& +a^{c,A}_{1}(\mathcal{B}_{b\overline 3})_{ij} H^{(c)}_{k}M^i_l ( \mathcal{B}_8^A)^{jkl}+ a^{c,A}_{2}(\mathcal{B}_{b\overline 3})_{ij} H^{(c)}_{k}M^i_l ( \mathcal{B}_8^A)^{jlk}+ a^{c,A}_{3}(\mathcal{B}_{b\overline 3})_{ij} H^{(c)}_{k}M^i_l (\mathcal{B}_8^A)^{klj} \nonumber\\
& + a^{c,A}_{4}(\mathcal{B}_{b\overline 3})_{ij} H^{(c)}_{k}M^k_l ( \mathcal{B}_8^A)^{ilj}+ a^{c,A}_{5}(\mathcal{B}_{b\overline 3})_{ij} H^{(c)}_{k}M^l_l ( \mathcal{B}_8^A)^{ikj}+ a^{c,A}_{6}(\mathcal{B}_{b\overline 3})_{ij} H^{(c)}_{k}M^k_l ( \mathcal{B}_8^A)^{ijl}\nonumber\\
& + a^{c,A}_{7}(\mathcal{B}_{b\overline 3})_{ij} H^{(c)}_{k}M^l_l ( \mathcal{B}_8^A)^{ijk}.
\end{align}
Each term in Eqs.~\eqref{amp1} and \eqref{amp2} can be interpreted as a topological diagram if the index contraction is understood as quark flowing.
The topological diagrams contributing to the $\mathcal{B}_{b\overline 3}\to \mathcal{B}^A_8 M$ decay are shown in Fig.~\ref{top7}.
The topological diagrams contributing to the $\mathcal{B}_{b\overline 3}\to \mathcal{B}^S_8 M$ decay can be obtained by replacing the antisymmetric quarks of $a_1$ to $a_{27}$ in Fig.~\ref{top7} with symmetric ones.
In Eqs.~\eqref{amp1} and \eqref{amp2}, we do not distinguish the tree and penguin contributions.
Note that the same contractions with different operators in the weak vertex are different topological amplitudes.
The light quark-loops in diagrams $a^{S,A}_{18-22}$, $a^{A}_{30,31}$ can be replaced by charm-loops; then, the $\mathcal{O}^{(c)}$ contributions are included.

The second-rank octet is given by
\begin{eqnarray}\label{B8}
 |\mathcal{B}_8\rangle=  \left( \begin{array}{ccc}
   \frac{1}{\sqrt 2} \Sigma^0+  \frac{1}{\sqrt 6} \Lambda^0    & \Sigma^+  & p \\
    \Sigma^- &   - \frac{1}{\sqrt 2} \Sigma^0+ \frac{1}{\sqrt 6} \Lambda^0   & n \\
    \Xi^- & \Xi^0 & -\sqrt{2/3}\Lambda^0 \\
  \end{array}\right).
\end{eqnarray}
The bottom baryon antitriplet can be expressed using the Levi-Civita tensor as
\begin{eqnarray}\label{sy2}
|\mathcal{B}_{b\overline 3}\rangle_{ij}=\epsilon_{ijk}|\mathcal{B}_{b\overline 3}\rangle^{k}\qquad {\rm with}\qquad |\mathcal{B}_{b\overline 3}\rangle^{k}=\left( \begin{array}{ccc}
     \Xi_b^- \\
    -\Xi_b^0  \\
    \Lambda_b^0 \\
  \end{array}\right).
\end{eqnarray}
The decay amplitude of the $\mathcal{B}_{b\overline 3}\to \mathcal{B}_8 M$ mode could be constructed by a (1,1)-rank baryon octet and first-rank charmed baryon antitriplet.
There are four and three covariant/contravariant indices in tensor $\langle (\mathcal{B}_{8})^{i}_j M^k_l |\mathcal{O}_{mn}^{(u,p)q}|\mathcal{B}_{b\overline 3}^{p}\rangle$ and $\langle (\mathcal{B}_{8})^{i}_j M^k_l |\mathcal{O}_{m}^{(c)}|\mathcal{B}_{b\overline 3}^{n}\rangle$, respectively.
The number of full contractions is $N=A^4_4=24$ and $N^c=A^3_3=6$.
Considering that $i\neq j$ in $(\mathcal{B}_{8})^{i}_j$, the number of terms contributing to the $\mathcal{B}_{b\overline 3}\to\mathcal{B}_8 M$ mode is $N+N^c = 18+4=22$.
The decay amplitude of the $\mathcal{B}_{b\overline 3}\to \mathcal{B}_8 M$ mode constructed by the second-rank octet tensors is
\begin{align}\label{amp3}
 \mathcal{A}(\mathcal{B}_{b\overline 3}\to \mathcal{B}_8M) = & A_{1}(\mathcal{B}_{b\overline 3})^{i} H^j_{kl}M^l_i( \mathcal{B}_8)^k_j
 +A_{2}(\mathcal{B}_{b\overline 3})^{i} H^j_{lk}M^l_j( \mathcal{B}_8)^k_i+ A_{3}(\mathcal{B}_{b\overline 3})^{i} H^j_{lk}M^l_i( \mathcal{B}_8)^k_j\nonumber\\
&+A_{4}(\mathcal{B}_{b\overline 3})^{i} H^j_{kl}M^l_j( \mathcal{B}_8)^k_i+ A_5(\mathcal{B}_{b\overline 3})^{i} H^j_{ik}M^l_j( \mathcal{B}_8)^k_l+ A_6(\mathcal{B}_{b\overline 3})^{i} H^j_{il}M^l_k( \mathcal{B}_8)^k_j\nonumber\\
&+A_7(\mathcal{B}_{b\overline 3})^{i} H^j_{ki}M^l_j( \mathcal{B}_8)^k_l  + A_8(\mathcal{B}_{b\overline 3})^{i} H^j_{li}M^l_k( \mathcal{B}_8)^k_j+
+ A_9(\mathcal{B}_{b\overline 3})^{i} H^j_{ik}M^l_l( \mathcal{B}_8)^k_j\nonumber\\
&+A_{10}(\mathcal{B}_{b\overline 3})^{i} H^j_{ki}M^l_l( \mathcal{B}_8)^k_j
+A_{11}(\mathcal{B}_{b\overline 3})^{i} H^j_{kj}M^l_i( \mathcal{B}_8)^k_l
+ A_{12}(\mathcal{B}_{b\overline 3})^{i} H^j_{lj}M^l_k( \mathcal{B}_8)^k_i\nonumber\\
& + A_{13}(\mathcal{B}_{c\overline 3})^{i} H^j_{kj}M^l_l( \mathcal{B}_8)^k_i
+A_{14}(\mathcal{B}_{b\overline 3})^{i} H^j_{ij}M^l_k( \mathcal{B}_8)^k_l
+ A_{15}(\mathcal{B}_{b\overline 3})^{i} H^j_{jl} M^l_k
(\mathcal{B}_8)^k_i\nonumber\\
&+A_{16}(\mathcal{B}_{b\overline 3})^{i} H^j_{jk}M^l_l( \mathcal{B}_8)^k_i
 + A_{17}(\mathcal{B}_{b\overline 3})^{i} H^j_{jk}M^l_i( \mathcal{B}_8)^k_l
 + A_{18}(\mathcal{B}_{b\overline 3})^{i} H^j_{ji}M^l_k( \mathcal{B}_8)^k_l\nonumber\\
&
 + A^c_{1}(\mathcal{B}_{b\overline 3})^{i} H^{(c)}_{j} M^j_k
(\mathcal{B}_8)^k_i+A^c_{2}(\mathcal{B}_{b\overline 3})^{i} H^{(c)}_{k}M^j_j( \mathcal{B}_8)^k_i
 + A^c_{3}(\mathcal{B}_{b\overline 3})^{i} H^{(c)}_{k}M^j_i( \mathcal{B}_8)^k_j
 \nonumber\\
&+ A^c_{4}(\mathcal{B}_{b\overline 3})^{i} H^{(c)}_{i}M^j_k( \mathcal{B}_8)^k_j.
\end{align}

By substituting Eqs.~\eqref{sy1} and \eqref{sy2} into each term of Eqs.~\eqref{amp1} and \eqref{amp2}, we can derive the relations between decay amplitudes constructed by (1,1)- and third-rank octets.
For example, the first term of Eq.~\eqref{amp1} is simplified as
\begin{align}
a^S_1(\mathcal{B}_{b\overline 3})_{ij} H^m_{kl}M^i_m ( \mathcal{B}_8^S)^{jkl} & = \frac{1}{\sqrt{6}}a^S_1\epsilon_{ijp}(\mathcal{B}_{b\overline 3})^{p} H^m_{kl}M^i_m \epsilon^{ljq}(\mathcal{B}_8)^k_q+\frac{1}{\sqrt{6}}a^S_1\epsilon_{ijp}(\mathcal{B}_{b\overline 3})^{p} H^m_{kl}M^i_m\epsilon^{lkq}(\mathcal{B}_8)^j_q\nonumber\\
& = -\frac{1}{\sqrt{6}}a^S_1\,(\mathcal{B}_{b\overline 3})^{i} H^j_{lk}M^l_j( \mathcal{B}_8)^k_i+\frac{2}{\sqrt{6}}a^S_1\,(\mathcal{B}_{b\overline 3})^{i} H^j_{kl}M^l_j( \mathcal{B}_8)^k_i\nonumber\\&~~~~~+\frac{1}{\sqrt{6}}a^S_1\,(\mathcal{B}_{b\overline 3})^{i} H^j_{ik}M^l_j( \mathcal{B}_8)^k_l-\frac{2}{\sqrt{6}}a^S_1\,(\mathcal{B}_{b\overline 3})^{i} H^j_{ki}M^l_j( \mathcal{B}_8)^k_l.
\end{align}
Thus, the diagram $a^S_1$ contributes to Eq.~\eqref{amp3} as
\begin{align}
&A_2 = -\frac{1}{\sqrt{6}}\,a^S_1+..., \qquad A_4 = \frac{2}{\sqrt{6}}\,a^S_1+...,\qquad A_5 = \frac{1}{\sqrt{6}}\,a^S_1+...,\qquad A_7 = -\frac{2}{\sqrt{6}}\,a^S_1+...\,\,.
\end{align}
The relations between the third- and (1,1)-rank topological amplitudes for the $\mathcal{B}_{b\overline 3}\to \mathcal{B}_8 M$ modes are derived as follows:
\begin{align}\label{sol}
 A_1  = &(a^A_6+a^A_{11}+a^A_{12})/\sqrt{2}-(a^S_4-a^S_{5}+2a^S_{10}-a^S_{11}
 -a^S_{12})/\sqrt{6}, \nonumber\\
 A_2  = & (-a^A_1+a^A_{3}-a^A_{5}-a^A_8-a^A_{9}+a^A_{14}+2a^A_{28})/\sqrt{2}-(a^S_1-2a^S_{2}+a^S_{3}-a^S_{4}
 +a^S_{6}-2a^S_7+a^S_{8}+a^S_{9}+3a^S_{14})/\sqrt{6}, \nonumber\\
 A_3  = &(a^A_5+a^A_{8}+a^A_{9})/\sqrt{2}-(a^S_4-a^S_{6}+2a^S_{7}-a^S_{8}
 -a^S_{9})/\sqrt{6}, \nonumber\\
 A_4  = &(-a^A_2-a^A_{3}-a^A_{6}-a^A_{11}-a^A_{12}+a^A_{13}+2a^A_{29})/\sqrt{2}
 -(-2a^S_1+a^S_{2}+a^S_{3}-a^S_{4}
 +a^S_{5}-2a^S_{10}+a^S_{11}+a^S_{12}
 +3a^S_{13})/\sqrt{6}, \nonumber\\
 A_5  = &(a^A_1-a^A_{3}+a^A_{5})/\sqrt{2}-(-a^S_1+2a^S_{2}-a^S_{3}+a^S_{4}
 -a^S_{6})/\sqrt{6}, \nonumber\\
 A_6  = &(-a^A_4+a^A_{10}-a^A_{12})/\sqrt{2}-(-a^S_5+a^S_{6}-a^S_{10}+2a^S_{11}
 -a^S_{12})/\sqrt{6}, \nonumber\\
 A_7  = &(a^A_2+a^A_{3}+a^A_{6})/\sqrt{2}-(2a^S_1-a^S_{2}-a^S_{3}+a^S_{4}
 -a^S_{5})/\sqrt{6}, \nonumber\\
 A_8  = &(a^A_4+a^A_{7}-a^A_{9})/\sqrt{2}-(a^S_5-a^S_{6}-a^S_{7}+2a^S_{8}
 -a^S_{9})/\sqrt{6}, \nonumber\\
 A_9  = &(-a^A_5+a^A_{15}-a^A_{17})/\sqrt{2}-(-a^S_4+a^S_{6}-a^S_{15}+2a^S_{16}
 -a^S_{17})/\sqrt{6}, \nonumber\\
 A_{10}  = &(-a^A_6+a^A_{16}+a^A_{17})/\sqrt{2}-(-a^S_4+a^S_{5}+2a^S_{15}-a^S_{16}
 -a^S_{17})/\sqrt{6}, \nonumber\\
 A_{11}  = &(-a^A_6+a^A_{19}+a^A_{20})/\sqrt{2}-(-a^S_4+a^S_{5}+2a^S_{18}-a^S_{19}
 -a^S_{20})/\sqrt{6}, \nonumber\\
 A_{12}  = &(-a^A_4-a^A_{7}+a^A_{9}-a^A_{18}+a^A_{20}+a^A_{21}+2a^A_{30})/\sqrt{2}-(-a^S_5+a^S_{6}+a^S_{7}-2a^S_{8}
 +a^S_{9}+a^S_{18}-2a^S_{19}
 +a^S_{20}+3a^S_{21})/\sqrt{6}, \nonumber\\
 A_{13}  = &(a^A_6-a^A_{16}-a^A_{17}-a^A_{19}-a^A_{20}+a^A_{22}+2a^A_{31})/\sqrt{2}-(a^S_4-a^S_{5}-2a^S_{15}+a^S_{16}
 +a^S_{17}-2a^S_{18}+a^S_{19}
 +a^S_{20}+3a^S_{22})/\sqrt{6}, \nonumber\\
 A_{14}  = &(a^A_4+a^A_{18}-a^A_{20})/\sqrt{2}-(a^S_5-a^S_{6}-a^S_{18}+2a^S_{19}
 -a^S_{20})/\sqrt{6}, \nonumber\\
 A_{15}  = &(a^A_4-a^A_{10}+a^A_{12}-a^A_{23}+a^A_{25}+a^A_{26}+2a^A_{32})/\sqrt{2}-(a^S_5-a^S_{6}+a^S_{10}-2a^S_{11}
 +a^S_{12}+a^S_{23}-2a^S_{24}
 +a^S_{25}+3a^S_{26})/\sqrt{6}, \nonumber\\
 A_{16}  = &(a^A_5-a^A_{15}+a^A_{17}-a^A_{24}-a^A_{25}+a^A_{27}+2a^A_{33})/\sqrt{2}-(a^S_4-a^S_{6}+a^S_{15}-2a^S_{16}
 +a^S_{17}-2a^S_{23}+a^S_{24}
 +a^S_{25}+3a^S_{27})/\sqrt{6}, \nonumber\\
 A_{17}  = &(-a^A_5+a^A_{24}+a^A_{25})/\sqrt{2}-(-a^S_4+a^S_{6}+2a^S_{23}-a^S_{24}
 -a^S_{25})/\sqrt{6}, \nonumber\\
 A_{18}  = &(-a^A_4+a^A_{23}-a^A_{25})/\sqrt{2}-(-a^S_5+a^S_{6}-a^S_{23}+2a^S_{24}
 -a^S_{25})/\sqrt{6}\nonumber\\
 A^c_{1}  = &(-a^{c,A}_{1}+a^{c,A}_{3}+a^A_{4}+2a^{c,A}_{6})/\sqrt{2}-(
 a^{c,S}_{1}-2a^{c,S}_{2}
 +a^{c,S}_{3}+3a^{c,S}_{4})/\sqrt{6}, \nonumber\\
 A^c_{2}  = &(-a^{c,A}_{2}-a^{c,A}_{3}+a^{c,A}_{5}+2a^{c,A}_{7})/\sqrt{2}-(
 -2a^{c,S}_{1}+a^{c,S}_{2}
 +a^{c,S}_{3}+3a^{c,S}_{5})/\sqrt{6}, \nonumber\\
 A^c_{3}  = &(a^{c,A}_{2}+a^{c,A}_{3})/\sqrt{2}-(2a^{c,S}_{1}-a^{c,S}_{2}
 -a^{c,S}_{3})/\sqrt{6}, \nonumber\\
 A^c_{4}  = &(a^{c,A}_{1}-a^{c,A}_{3})/\sqrt{2}-(-a^{c,S}_{1}+2a^{c,S}_{2}
 -a^{c,S}_{3})/\sqrt{6}.
\end{align}
According to Eq.~\eqref{sol}, the topological diagrams of the $\mathcal{B}_{b\overline 3}\to \mathcal{B}^S_{8}M$ and $\mathcal{B}_{b\overline 3}\to \mathcal{B}^A_{8}M$ transitions are not independent.
If the topological diagrams of the $\mathcal{B}_{b\overline 3}\to \mathcal{B}^S_{8}M$ mode are known, the topological diagrams of the $\mathcal{B}_{b\overline 3}\to \mathcal{B}^A_{8}M$ mode are determined, and vice versa.

Operator $\mathcal{O}_{ij}^k$ is a reducible representation of the $SU(3)$ group.
It can be decomposed as $3 \otimes\overline 3 \otimes 3 =  3_p\oplus3_t\oplus \overline 6 \oplus 15$.
Then, the coefficient matrix $H$ can be written as
\begin{align}\label{c8}
  H^k_{ij}= &\frac{1}{8}H(15)^k_{ij}+\frac{1}{4}\epsilon_{ijl}H(\overline 6)^{lk}+\delta_j^k\Big(\frac{3}{8}H( 3_t)_i-\frac{1}{8}H(3_p)_i\Big)+
  \delta_i^k\Big(\frac{3}{8}H( 3_p)_j-\frac{1}{8}H( 3_t)_j\Big),
\end{align}
in which $H( 3_p)_i = H_{li}^l$ and $H( 3_t)_i = H_{il}^l$.
The $SU(3)$ irreducible amplitude of the $\mathcal{B}_{b\overline 3}\to \mathcal{B}_8M$ decay is constructed as
\begin{align}\label{amp4}
{\cal A}^{\rm IR}(\mathcal{B}_{b\overline 3}\to \mathcal{B}_8M)& =
b_1(\mathcal{B}_{b\overline 3})^i H(\overline 6)_{ij}^kM^j_l( \mathcal{B}_8)^l_k +b_2(\mathcal{B}_{b\overline 3})^i H(\overline 6)_{ij}^kM^l_k(\mathcal{B}_8)^j_l+ b_3(\mathcal{B}_{b\overline 3})^i H(\overline 6)_{ij}^kM^l_l(\mathcal{B}_8)^j_k\nonumber\\ &  +b_4(\mathcal{B}_{b\overline 3})^i H(\overline 6)_{jl}^kM^j_i(\mathcal{B}_8)^l_k +b_5(\mathcal{B}_{b\overline 3})^i H(\overline 6)_{jl}^kM^l_k(\mathcal{B}_8)^j_i + b_{6}(\mathcal{B}_{b\overline 3})^i H({15})_{ij}^kM^j_l(\mathcal{B}_8)^l_k\nonumber\\
  & + b_{7}(\mathcal{B}_{b\overline 3})^i H({15})_{ij}^kM^l_k(\mathcal{B}_8)^j_l
  +b_{8}(\mathcal{B}_{b\overline 3})^i H({15})_{ij}^kM^l_l(\mathcal{B}_8)^j_k  + b_{9}(\mathcal{B}_{b\overline 3})^i H({15})_{jl}^kM^j_i(\mathcal{B}_8)^l_k \nonumber\\&+ b_{10}(\mathcal{B}_{b\overline 3})^i H({15})_{jl}^kM^l_k(\mathcal{B}_8)^j_i
+b_{11} (\mathcal{B}_{b\overline 3})^i H(3_p)_i M_k^j(\mathcal{B}_8)_j^k
  +b_{12} (\mathcal{B}_{b\overline 3})^i H(3_p)_k M_j^j(\mathcal{B}_8)_i^k\nonumber\\ &+b_{13} (\mathcal{B}_{b\overline 3})^i H(3_p)_k M_i^j(\mathcal{B}_8)_j^k +b_{14} (\mathcal{B}_{b\overline 3})^i H(3_p)_k M_j^k(\mathcal{B}_8)_i^j
+b_{15} (\mathcal{B}_{b\overline 3})^i H(3_t)_i M_k^j(\mathcal{B}_8)_j^k
\nonumber\\  & +b_{16} (\mathcal{B}_{b\overline 3})^i H(3_t)_k M_j^j(\mathcal{B}_8)_i^k +b_{17} (\mathcal{B}_{b\overline 3})^i H(3_t)_k M_i^j(\mathcal{B}_8)_j^k +b_{18} (\mathcal{B}_{b\overline 3})^i H(3_t)_k M_j^k(\mathcal{B}_8)_i^j\nonumber\\
&
 + b^c_{1}(\mathcal{B}_{b\overline 3})^{i} H(3)^{(c)}_{j} M^j_k
(\mathcal{B}_8)^k_i+b^c_{2}(\mathcal{B}_{b\overline 3})^{i} H(3)^{(c)}_{k}M^j_j( \mathcal{B}_8)^k_i
 + b^c_{3}(\mathcal{B}_{b\overline 3})^{i} H(3)^{(c)}_{k}M^j_i( \mathcal{B}_8)^k_j
 \nonumber\\
&+ b^c_{4}(\mathcal{B}_{b\overline 3})^{i} H(3)^{(c)}_{i}M^j_k( \mathcal{B}_8)^k_j.
\end{align}
By substituting Eq.~\eqref{c8} into each term of Eq.~\eqref{amp3},
the relations between Eqs.~\eqref{amp3} and \eqref{amp4} are derived as
\begin{align}\label{sol2}
 &b_1  =\frac{A_6-A_8}{4}, \qquad b_2  =\frac{A_5-A_7}{4},  \qquad b_3 = \frac{A_9-A_{10}}{4}, \qquad b_4 = \frac{-A_1+A_3}{4}, \qquad b_5 = \frac{-A_2+A_4}{4}, \nonumber\\ &b_{6} =\frac{A_6+A_8}{8}, \qquad
  b_{7}  =\frac{A_5+A_7}{8},  \qquad b_{8} = \frac{A_9+A_{10}}{8}, \qquad b_{9} = \frac{A_1+A_3}{8},\qquad  b_{10} = \frac{A_2+A_4}{8},\nonumber\\
& b_{11}= -\frac{1}{8}A_5+\frac{3}{8}A_7-\frac{1}{8}A_6+\frac{3}{8}A_8+ A_{18},\qquad b_{15} =\frac{3}{8}A_5-\frac{1}{8}A_7+\frac{3}{8}A_6-\frac{1}{8}A_8+A_{14},
\nonumber\\
 & b_{12} = \frac{3}{8}A_2-\frac{1}{8}A_4+\frac{3}{8}A_9 - \frac{1}{8}A_{10}+A_{16}, \qquad  b_{16} = -\frac{1}{8}A_2+\frac{3}{8}A_4-\frac{1}{8}A_9 + \frac{3}{8}A_{10}+A_{13},
   \nonumber\\
& b_{13} = -\frac{1}{8}A_1+\frac{3}{8}A_3+\frac{3}{8}A_5 - \frac{1}{8}A_7+A_{17},\qquad   b_{17} = \frac{3}{8}A_1-\frac{1}{8}A_3-\frac{1}{8}A_5 + \frac{3}{8}A_7+A_{11},
 \nonumber\\
& b_{14} = -\frac{1}{8}A_2+\frac{3}{8}A_4+\frac{3}{8}A_6 - \frac{1}{8}A_8+A_{15},\qquad   b_{18}  = \frac{3}{8}A_2-\frac{1}{8}A_4-\frac{1}{8}A_6 + \frac{3}{8}A_8+A_{12},\nonumber\\
 &
b^c_1= A^c_1,\qquad b^c_2= A^c_2,\qquad b^c_3= A^c_3,\qquad b^c_4= A^c_4.
\end{align}
The inverse solution of Eq.~\eqref{sol2} is
\begin{align}\label{sol3}
 &A_1  =-2b_4+4b_9, \qquad A_2  =-2b_5+4b_{10},  \qquad A_3  =2b_4+4b_{9}, \qquad A_4  =2b_5+4b_{10}, \qquad A_5  =2b_2+4b_{7}, \nonumber\\  &A_6  =2b_1+4b_6, \qquad A_7  =-2b_2+4b_{7},  \qquad A_8  =-2b_1+4b_{6}, \qquad A_9  =2b_3+4b_{8}, \qquad A_{10}  =-2b_3+4b_{8}, \nonumber\\
&A_{11}  =b_{17}+b_2+b_4-b_7-b_9, \qquad A_{12}  =b_{18}+b_1+b_5-b_6-b_{10}, \qquad A_{13}  =b_{16}+b_3-b_5-b_8-b_{10}, \nonumber\\
&A_{14}  =b_{15}-b_1-b_2-b_6-b_7, \qquad A_{15}  =b_{14}-b_1-b_5-b_6-b_{10}, \qquad A_{16}  =b_{12}-b_3+b_5-b_8-b_{10}, \nonumber\\
&A_{17}  =b_{13}-b_2-b_4-b_7-b_9, \qquad A_{18}  =b_{11}+b_1+b_2-b_6-b_7
\nonumber\\
 &
A^c_1 = b^c_1,\qquad A^c_2 = b^c_2,\qquad A^c_3 = b^c_3,\qquad A^c_4 = b^c_4.
\end{align}
By substituting Eq.~\eqref{sol} into Eq.~\eqref{sol2}, we obtain the equations for $SU(3)$ irreducible amplitudes decomposed in terms of topological amplitudes,
\begin{align}\label{sol11}
b_1& =-\sqrt{2}(2 a^A_4+a^A_7-a^A_9-a^A_{10}+a^A_{12})/8+\sqrt{6}(2 a^S_{5}-2 a^S_6-a^S_7+2 a^S_8-a^S_9+a^S_{10}-2 a^S_{11}+a^S_{12})/24,\nonumber\\
b_2&=\sqrt{2}(a^A_1-a^A_2-2 a^A_3+a^A_5-a^A_6)/8+\sqrt{6}(3 a^S_1-3 a^S_2-a^S_5+a^S_6)/24,\nonumber\\
b_3&=-\sqrt{2}(a^A_5-a^A_6-a^A_{15}+a^A_{16}+2 a^A_{17})/8+\sqrt{6}(a^S_5-a^S_6+3 a^S_{15}-3 a^S_{16})/24,\nonumber\\
b_4&=\sqrt{2}(a^A_5-a^A_6+a^A_8+a^A_9-a^A_{11}-a^A_{12})/8
-\sqrt{6}(a^S_5-a^S_6+2 a^S_7-a^S_8-a^S_9-2 a^S_{10}+a^S_{11}+a^S_{12})/24,\nonumber\\
b_5&=\sqrt{2}(a^A_1-a^A_2-2 a^A_3+a^A_5-a^A_6+a^A_8+a^A_9-a^A_{11}-a^A_{12}+a^A_{13}-a^A_{14}-2 a^A_{28}+2 a^A_{29})/8\nonumber\\&~~~~~~+\sqrt{6}(3 a^S_1-3 a^S_2-a^S_5+a^S_6-2 a^S_7+a^S_8+a^S_9+2 a^S_{10}-a^S_{11}-a^S_{12}-3 a^S_{13}+3 a^S_{14})/24,\nonumber\\
b_6&=\sqrt{2}(a^A_7-a^A_9+a^A_{10}-a^A_{12})/16+\sqrt{6}(a^S_7-2 a^S_8+a^S_9+a^S_{10}-2 a^S_{11}+a^S_{12})/48,\nonumber\\
b_7&=(a^A_1+a^A_2+a^A_5+a^A_6)/16-\sqrt{6}(a^S_1+a^S_2-2 a^S_3+2 a^S_4-a^S_5-a^S_6)/48, \nonumber\\
b_8&=-\sqrt{2}(a^A_5+a^A_6-a^A_{15}-a^A_{16})/16+\sqrt{6}(2 a^S_4-a^S_5-a^S_6-a^S_{15}-a^S_{16}+2 a^S_{17})/48, \nonumber\\
b_9&=\sqrt{2}(a^A_5+a^A_6+a^A_8+a^A_9+a^A_{11}+a^A_{12})/16+\sqrt{6}(-2 a^S_4+a^S_5+a^S_6-2 a^S_7+a^S_8+a^S_9-2 a^S_{10}+a^S_{11}+a^S_{12})/48,\nonumber\\
b_{10}&=-\sqrt{2}(a^A_1+a^A_2+a^A_5+a^A_6+a^A_8+a^A_9+a^A_{11}+a^A_{12}
-a^A_{13}-a^A_{14}-2 a^A_{28}-2 a^A_{29})/16\nonumber\\&~~~~~~+\sqrt{6}(a^S_1+a^S_2-2 a^S_3+2 a^S_4-a^S_5-a^S_6+2 a^S_7-a^S_8-a^S_9+2 a^S_{10}-a^S_{11}-a^S_{12}-3 a^S_{13}-3 a^S_{14})/48,\nonumber\\
b_{11}&=-\sqrt{2}(a^A_1-3 a^A_2-4 a^A_3+4 a^A_4+a^A_5-3 a^A_6-3a^A_7+3a^A_9+a^A_{10}-a^A_{12}-8 a^A_{23}+8 a^A_{25})/16\nonumber\\&~~~~~~-\sqrt{6}(7 a^S_1-5 a^S_2-2 a^S_3+2 a^S_4-7 a^S_5+5 a^S_6-3a^S_7+6 a^S_8-3a^S_9+a^S_{10}-2 a^S_{11}+a^S_{12}-8 a^S_{23}+16 a^S_{24}-8 a^S_{25})/48,\nonumber\\
b_{12}&=-\sqrt{2}(3 a^A_1-a^A_2-4 a^A_3-2 a^A_5-2 a^A_6+3 a^A_8+3a^A_9-a^A_{11}-a^A_{12}+a^A_{13}-3 a^A_{14}+5 a^A_{15}+a^A_{16}-4 a^A_{17}\nonumber\\&~~~~~~+8 a^A_{24}+8 a^A_{25}-8 a^A_{27}-6 a^A_{28}+2 a^A_{29}-16 a^A_{33}) /16-\sqrt{6}(5 a^S_1-7 a^S_2+2 a^S_3+4 a^S_4-2 a^S_5-2 a^S_6-6 a^S_7+3 a^S_8\nonumber\\&~~~~~~~~+3a^S_9+2 a^S_{10}-a^S_{11}-a^S_{12}-3 a^S_{13}+9 a^S_{14}+3 a^S_{15}-9 a^S_{16}+6 a^S_{17}-16 a^S_{23}+8 a^S_{24}+8 a^S_{25}+24 a^S_{27})/48,\nonumber\\
b_{13}&=\sqrt{2}(3 a^A_1-a^A_2-4 a^A_3-2 a^A_5-2 a^A_6+3 a^A_8+3a^A_9-a^A_{11}-a^A_{12}+8 a^A_{24}+8 a^A_{25})/16\nonumber\\&~~~~~~+\sqrt{6}(5 a^S_1-7 a^S_2+2 a^S_3+4 a^S_4-2 a^S_5-2 a^S_6-6 a^S_7+3a^S_8+3a^S_9+2 a^S_{10}-a^S_{11}-a^S_{12}-16 a^S_{23}+8 a^S_{24}+8 a^S_{25})/48,\nonumber\\
b_{14}&=\sqrt{2}(a^A_1-3 a^A_2-4 a^A_3+4 a^A_4+a^A_5-3 a^A_6-a^A_7+a^A_8+2 a^A_9-5 a^A_{10}-3 a^A_{11}+2 a^A_{12}+3 a^A_{13}-a^A_{14}-8 a^A_{23}+8 a^A_{25}\nonumber\\&~~~~~~+8 a^A_{26}-2 a^A_{28}+6 a^A_{29}+16 a^A_{32})/16+\sqrt{6}(7 a^S_1-5 a^S_2-2 a^S_3+2 a^S_4-7 a^S_5+5 a^S_6-3 a^S_7+3 a^S_8+a^S_{10}+7 a^S_{11}\nonumber\\&~~~~~~~~-8 a^S_{12}-9 a^S_{13}+3 a^S_{14}-8 a^S_{23}+16 a^S_{24}-8 a^S_{25}-24 a^S_{26})/48,\nonumber\\
b_{15}&=\sqrt{2}(3 a^A_1-a^A_2-4 a^A_3+4 a^A_4+3 a^A_5-a^A_6-a^A_7+a^A_9+3 a^A_{10}-3 a^A_{12}+8 a^A_{18}-8 a^A_{20})/16\nonumber\\&~~~~~~+\sqrt{6}(5 a^S_1-7 a^S_2+2 a^S_3-2 a^S_4-5 a^S_5+7 a^S_6-a^S_7+2 a^S_8-a^S_9+3 a^S_{10}-6 a^S_{11}+3 a^S_{12}+8 a^S_{18}-16 a^S_{19}+8 a^S_{20})/48,\nonumber\\
b_{16}&=\sqrt{2}(a^A_1-3 a^A_2-4 a^A_3+2 a^A_5+2 a^A_6+a^A_8+a^A_9-3 a^A_{11}-3 a^A_{12}+3 a^A_{13}-a^A_{14}-a^A_{15}-5 a^A_{16}-4 a^A_{17}-8 a^A_{19}-8 a^A_{20}\nonumber\\&~~~~~~+8 a^A_{22}-2 a^A_{28}+6 a^A_{29}+16 a^A_{31})/16+\sqrt{6}(7 a^S_1-5 a^S_2-2 a^S_3-4 a^S_4+2 a^S_5+2 a^S_6-2 a^S_7+a^S_8+a^S_9+6 a^S_{10}\nonumber\\&~~~~~~~~-3 a^S_{11}-3 a^S_{12}-9 a^S_{13}+3 a^S_{14}+9 a^S_{15}-3 a^S_{16}-6 a^S_{17}+16 a^S_{18}-8 a^S_{19}-8 a^S_{20}-24 a^S_{22})/48,\nonumber\\
b_{17}&=-\sqrt{2}(a^A_1-3 a^A_2-4 a^A_3+2 a^A_5+2 a^A_6+a^A_8+a^A_9-3 a^A_{11}-3 a^A_{12}-8 a^A_{19}-8 a^A_{20})/16\nonumber\\&~~~~~~-\sqrt{6}(7 a^S_1-5 a^S_2-2 a^S_3-4 a^S_4+2 a^S_5+2 a^S_6-2 a^S_7+a^S_8+a^S_9+6 a^S_{10}-3 a^S_{11}-3 a^S_{12}+16 a^S_{18}-8 a^S_{19}-8 a^S_{20})/48,\nonumber\\
b_{18}&=-\sqrt{2}(3 a^A_1-a^A_2-4 a^A_3+4 a^A_4+3 a^A_5-a^A_6+5 a^A_7+3 a^A_8-2 a^A_9+a^A_{10}-a^A_{11}-2 a^A_{12}+a^A_{13}-3 a^A_{14}+8 a^A_{18}-8 a^A_{20}\nonumber\\&~~~~~~-8 a^A_{21}-6 a^A_{28}+2 a^A_{29}-16 a^A_{30})/16-\sqrt{6}(5 a^S_1-7 a^S_2+2 a^S_3-2 a^S_4-5 a^S_5+7 a^S_6-a^S_7-7 a^S_8+8 a^S_9+3 a^S_{10}\nonumber\\&~~~~~~~~-3 a^S_{11}-3 a^S_{13}+9 a^S_{14}+8 a^S_{18}-16 a^S_{19}+8 a^S_{20}+24 a^S_{21})/48,
\nonumber\\
 b^c_{1}  &= (-a^{c,A}_{1}+a^{c,A}_{3}+a^A_{4}+2a^{c,A}_{6})/\sqrt{2}-(
 a^{c,S}_{1}-2a^{c,S}_{2}
 +a^{c,S}_{3}+3a^{c,S}_{4})/\sqrt{6}, \nonumber\\
 b^c_{2}  &= (-a^{c,A}_{2}-a^{c,A}_{3}+a^{c,A}_{5}+2a^{c,A}_{7})/\sqrt{2}-(
 -2a^{c,S}_{1}+a^{c,S}_{2}
 +a^{c,S}_{3}+3a^{c,S}_{5})/\sqrt{6}, \nonumber\\
 b^c_{3} & = (a^{c,A}_{2}+a^{c,A}_{3})/\sqrt{2}-(2a^{c,S}_{1}-a^{c,S}_{2}
 -a^{c,S}_{3})/\sqrt{6}, \nonumber\\
 b^c_{4}  &= (a^{c,A}_{1}-a^{c,A}_{3})/\sqrt{2}-(-a^{c,S}_{1}+2a^{c,S}_{2}
 -a^{c,S}_{3})/\sqrt{6}.
\end{align}
Note that the inverse solution does not exist because the number of terms of topological amplitudes is larger than the $SU(3)$ irreducible amplitudes.

The terms constructed by $H(\overline 6)_{ij}^k$ in Eq.~\eqref{amp4} can be written as
\begin{align}\label{c1}
 b_1(\mathcal{B}_{b\overline 3})^i H(\overline 6)_{ij}^kM^j_l( \mathcal{B}_8)^l_k &= b_1(\mathcal{B}_{b\overline 3})_{ji} H(\overline 6)^{ki}M^j_l( \mathcal{B}_8)^l_k,\nonumber\\
b_2(\mathcal{B}_{b\overline 3})^i H(\overline 6)_{ij}^kM^l_k(\mathcal{B}_8)^j_l &= b_2(\mathcal{B}_{b\overline 3})_{ji} H(\overline 6)^{ki}M^l_k( \mathcal{B}_8)^j_l,\nonumber\\
b_3(\mathcal{B}_{b\overline 3})^i H(\overline 6)_{ij}^kM^l_l(\mathcal{B}_8)^j_k&=b_3(\mathcal{B}_{b\overline 3})_{ji} H(\overline 6)^{ki}M^l_l( \mathcal{B}_8)^j_k,\nonumber\\
b_4(\mathcal{B}_{b\overline 3})^i H(\overline 6)_{jl}^kM^j_i(\mathcal{B}_8)^l_k&=b_4\big[(\mathcal{B}_{b\overline 3})_{jl} H(\overline 6)^{ki}M^j_i( \mathcal{B}_8)^l_k-(\mathcal{B}_{b\overline 3})_{ji} H(\overline 6)^{ki}M^j_l( \mathcal{B}_8)^l_k\nonumber\\&~~~~~+(\mathcal{B}_{b\overline 3})_{ji} H(\overline 6)^{ki}M^l_l( \mathcal{B}_8)^j_k\big],\nonumber\\
 b_5(\mathcal{B}_{b\overline 3})^i H(\overline 6)_{jl}^kM^l_k(\mathcal{B}_8)^j_i&=-b_5\big[(\mathcal{B}_{b\overline 3})_{ji} H(\overline 6)^{ki}M^l_k( \mathcal{B}_8)^j_l+(\mathcal{B}_{b\overline 3})_{jl} H(\overline 6)^{ki}M^j_i( \mathcal{B}_8)^l_k\big],
\end{align}
 where $(\mathcal{B}_{b\overline 3})^i=\epsilon^{ijk}(\mathcal{B}_{b\overline 3})_{jk}/2$
is used.
There are four invariant tensors in the RHS of Eq.~\eqref{c1}.
It indicates that the degree of freedom of the $SU(3)$ irreducible amplitudes associated with the $\overline 6$ representation is four.
We can define four new parameters to replace $b_1\sim b_5$,
\begin{align}\label{c9}
  b_1^\prime = b_1 - b_4,\qquad b_2^\prime = b_2 - b_5,\qquad b_3^\prime = b_3 +b_4,\qquad b_4^\prime = b_4 - b_5.
\end{align}

The amplitude of the $\mathcal{B}_{b\overline 3}\to \mathcal{B}_8M$ decay can also be constructed by third-rank octet tensors and $SU(3)$ irreducible operators.
The $SU(3)$ irreducible amplitude of the $\mathcal{B}_{b\overline 3}\to \mathcal{B}_8^SM$ decay constructed by third-rank octet tensors is
\begin{align}\label{IS}
 \mathcal{A}^S(\mathcal{B}_{b\overline 3}\to \mathcal{B}_8^SM) & = c^S_1(\mathcal{B}_{b\overline 3})_{ij} H(15)^m_{kl}M^i_m (\mathcal{B}_8^S)^{jkl} + c^S_2(\mathcal{B}_{b\overline 3})_{ij}H(\overline 6)^m_{kl}M^i_m (\mathcal{B}_8^S)^{jkl}+ c^S_3(\mathcal{B}_{b\overline 3})_{ij}H(15)^m_{kl}M^i_m (\mathcal{B}_8^S)^{klj} \nonumber\\
   & + c^S_4(\mathcal{B}_{b\overline 3})_{ij} H(15)^i_{kl}M^j_m (\mathcal{B}_8^S)^{klm}  + c^S_5(\mathcal{B}_{b\overline 3})_{ij} H(15)^i_{kl}M^j_m (\mathcal{B}_8^S)^{kml}  + c^S_6(\mathcal{B}_{b\overline 3})_{ij} H(\overline 6)^i_{kl}M^j_m (\mathcal{B}_8^S)^{kml} \nonumber\\
   &+ c^S_7(\mathcal{B}_{b\overline 3})_{ij} H(15)^i_{kl}M^k_m (\mathcal{B}_8^S)^{jlm}+ c^S_8(\mathcal{B}_{b\overline 3})_{ij} H(15)^i_{kl}M^k_m (\mathcal{B}_8^S)^{jml} + c^S_9(\mathcal{B}_{b\overline 3})_{ij} H(15)^i_{kl}M^k_m (\mathcal{B}_8^S)^{lmj} \nonumber\\
 &+ c^S_{10}(\mathcal{B}_{b\overline 3})_{ij} H(\overline 6)^i_{kl}M^k_m (\mathcal{B}_8^S)^{jlm}  + c^S_{11}(\mathcal{B}_{b\overline 3})_{ij} H(\overline 6)^i_{kl}M^k_m (\mathcal{B}_8^S)^{jml} + c^S_{12}(\mathcal{B}_{b\overline 3})_{ij} H(\overline 6)^i_{kl}M^k_m (\mathcal{B}_8^S)^{lmj} \nonumber\\
&+ c^S_{13}(\mathcal{B}_{b\overline 3})_{ij} H(15)^m_{kl}M^l_m (\mathcal{B}_8^S)^{ikj}+ c^S_{14}(\mathcal{B}_{b\overline 3})_{ij} H(\overline 6)^m_{kl}M^l_m (\mathcal{B}_8^S)^{ikj} + c^S_{15}(\mathcal{B}_{b\overline 3})_{ij} H(15)^i_{kl}M^m_m (\mathcal{B}_8^S)^{jkl} \nonumber\\
 &+ c^S_{16}(\mathcal{B}_{c\overline 3})_{ij} H(\overline 6)^i_{kl}M^m_m (\mathcal{B}_8^S)^{jkl} + c^S_{17}(\mathcal{B}_{b\overline 3})_{ij} H(15)^i_{kl}M^m_m (\mathcal{B}_8^S)^{klj}  + c^S_{18}(\mathcal{B}_{b\overline 3})_{ij} H(3_t)_{k}M^i_m (\mathcal{B}_8^S)^{jkm}\nonumber\\
 & + c^S_{19}(\mathcal{B}_{b\overline 3})_{ij} H(3_t)_{k}M^i_m (\mathcal{B}_8^S)^{jmk} + c^S_{20}(\mathcal{B}_{b\overline 3})_{ij} H(3_t)_{k}M^i_m (\mathcal{B}_8^S)^{kmj} + c^S_{21}(\mathcal{B}_{b\overline 3})_{ij} H(3_t)_{k}M^k_m (\mathcal{B}_8^S)^{imj}\nonumber\\
& + c^S_{22}(\mathcal{B}_{c\overline 3})_{ij} H(3_t)_{k}M^m_m (\mathcal{B}_8^S)^{ikj} + c^S_{23}(\mathcal{B}_{b\overline 3})_{ij} H(3_p)_{k}M^i_m (\mathcal{B}_8^S)^{jkm}+ c^S_{24}(\mathcal{B}_{b\overline 3})_{ij} H(3_p)_{k}M^i_m (\mathcal{B}_8^S)^{jmk}\nonumber\\
& + c^S_{25}(\mathcal{B}_{b\overline 3})_{ij} H(3_p)_{k}M^i_m (\mathcal{B}_8^S)^{kmj} + c^S_{26}(\mathcal{B}_{b\overline 3})_{ij} H(3_p)_{k}M^k_m (\mathcal{B}_8^S)^{imj}+ c^S_{27}(\mathcal{B}_{b\overline 3})_{ij} H(3_p)_{k}M^m_m (\mathcal{B}_8^S)^{ikj}\nonumber\\& +c^{c,S}_{1}(\mathcal{B}_{b\overline 3})_{ij} H(3)^c_{k}M^i_l ( \mathcal{B}_8^S)^{jkl}+ c^{c,S}_{2}(\mathcal{B}_{b\overline 3})_{ij} H(3)^c_{k}M^i_l ( \mathcal{B}_8^S)^{jlk}+ c^{c,S}_{3}(\mathcal{B}_{b\overline 3})_{ij} H(3)^c_{k}M^i_l (\mathcal{B}_8^S)^{klj} \nonumber\\
& + c^{c,S}_{4}(\mathcal{B}_{b\overline 3})_{ij} H(3)^c_{k}M^k_l ( \mathcal{B}_8^S)^{ilj}+ c^{c,S}_{5}(\mathcal{B}_{c\overline 3})_{ij} H(3)^c_{k}M^l_l ( \mathcal{B}_8^S)^{ikj}.
\end{align}
The relations of the $SU(3)$ irreducible amplitudes and the topological amplitudes are derived as
\begin{align}\label{sol6}
   & c^S_1 = \frac{a^S_1+a^S_2}{8},\quad c^S_2 = \frac{a^S_1-a^S_2}{4},  \quad c^S_3 = \frac{a^S_3}{8},  \quad c^S_4 = \frac{a^S_4}{8}, \quad c^S_5 = \frac{a^S_5+a^S_6}{8}, \quad  c^S_6 = \frac{a^S_5-a^S_6}{4},\nonumber\\
   & c^S_7 = \frac{a^S_7+a^S_{10}}{8},\quad c^S_{10} = \frac{a^S_7-a^S_{10}}{4},  \quad c^S_8 = \frac{a^S_8 + a^S_{11}}{8},  \quad c^S_{11} = \frac{a^S_8 - a^S_{11}}{4}, \quad c^S_9 = \frac{a^S_9+a^S_{12}}{8}, \quad  c^S_{12} = \frac{a^S_9-a^S_{12}}{4},\nonumber\\
   & c^S_{13} = \frac{a^S_{13}+a^S_{14}}{8},\quad c^S_{14} = \frac{a^S_{13}-a^S_{14}}{4},  \quad c^S_{15} = \frac{a^S_{15} + a^S_{16}}{8},  \quad c^S_{16} = \frac{a^S_{15} - a^S_{16}}{4}, \quad c^S_{17} = \frac{a^S_{17}}{8},\nonumber\\
   &  c^S_{18} = \frac{3}{8}a^S_1- \frac{1}{8}a^S_2 - \frac{1}{4}a^S_4- \frac{1}{8}a^S_7 +\frac{3}{8}a^S_{10}+ a^S_{18},\quad c^S_{23} = -\frac{1}{8}a^S_1+ \frac{3}{8}a^S_2 - \frac{1}{4}a^S_4+ \frac{3}{8}a^S_7 -\frac{1}{8}a^S_{10}+ a^S_{23},\nonumber\\
   &  c^S_{19} = -\frac{1}{8}a^S_1+ \frac{3}{8}a^S_2 + \frac{1}{8}a^S_{5}- \frac{3}{8}a^S_{6} -\frac{1}{8}a^S_{8}+ \frac{3}{8}a^S_{11}+ a^S_{19},\quad c^S_{24} = \frac{3}{8}a^S_1- \frac{1}{8}a^S_2 - \frac{3}{8}a^S_{5}+ \frac{1}{8}a^S_{6} +\frac{3}{8}a^S_{8}- \frac{1}{8}a^S_{11}+ a^S_{24},\nonumber\\
   &  c^S_{20} = \frac{1}{4}a^S_3- \frac{3}{8}a^S_5 + \frac{1}{8}a^S_{6}- \frac{1}{8}a^S_{9} +\frac{3}{8}a^S_{12}+ a^S_{20},\quad c^S_{25} = \frac{1}{4}a^S_3+ \frac{1}{8}a^S_5 - \frac{3}{8}a^S_{6}+ \frac{3}{8}a^S_{9} -\frac{1}{8}a^S_{12}+ a^S_{25},
\nonumber\\
   &  c^S_{21} = -\frac{3}{8}a^S_8+ \frac{1}{8}a^S_{11} + \frac{3}{8}a^S_{9}- \frac{1}{8}a^S_{12} -\frac{1}{8}a^S_{13} + \frac{3}{8}a^S_{14}+a^S_{21},\quad  c^S_{26} = \frac{1}{8}a^S_8- \frac{3}{8}a^S_{11} - \frac{1}{8}a^S_{9}+\frac{3}{8}a^S_{12} +\frac{3}{8}a^S_{13} - \frac{1}{8}a^S_{14}+a^S_{26},
\nonumber\\
   &  c^S_{22} = \frac{3}{8}a^S_{13}- \frac{1}{8}a^S_{14} -\frac{3}{8}a^S_{15}+ \frac{1}{8}a^S_{16} +\frac{1}{4}a^S_{17} +a^S_{22},\quad  c^S_{27} = -\frac{1}{8}a^S_{13}+\frac{3}{8}a^S_{14} +\frac{1}{8}a^S_{15}- \frac{3}{8}a^S_{16} +\frac{1}{4}a^S_{17} +a^S_{27},\nonumber\\
& c^{c,S}_1 =  a^{c,S}_1,\qquad c^{c,S}_2 =  a^{c,S}_2,\qquad c^{c,S}_3 =  a^{c,S}_3,\qquad c^{c,S}_4 =  a^{c,S}_4,\qquad  c^{c,S}_5 =  a^{c,S}_5.
\end{align}
The $SU(3)$ irreducible amplitude of the $\mathcal{B}_{c\overline 3}\to \mathcal{B}_8^AM$ decay constructed by third-rank octet tensors is
\begin{align}\label{IA}
 \mathcal{A}^A(\mathcal{B}_{c\overline 3}\to \mathcal{B}_8^AM) & = c^A_1(\mathcal{B}_{c\overline 3})_{ij} H(15)^m_{kl}M^i_m (\mathcal{B}_8^A)^{jkl} + c^A_2(\mathcal{B}_{c\overline 3})_{ij}H(\overline 6)^m_{kl}M^i_m (\mathcal{B}_8^A)^{jkl}+ c^A_3(\mathcal{B}_{c\overline 3})_{ij}H(\overline 6)^m_{kl}M^i_m (\mathcal{B}_8^A)^{klj} \nonumber\\
   & + c^A_4(\mathcal{B}_{c\overline 3})_{ij} H(\overline 6)^i_{kl}M^j_m (\mathcal{B}_8^A)^{klm}  + c^A_5(\mathcal{B}_{c\overline 3})_{ij} H(15)^i_{kl}M^j_m (\mathcal{B}_8^A)^{kml}  + c^A_6(\mathcal{B}_{c\overline 3})_{ij} H(\overline 6)^i_{kl}M^j_m (\mathcal{B}_8^A)^{kml} \nonumber\\
   &+ c^A_7(\mathcal{B}_{c\overline 3})_{ij} H(15)^i_{kl}M^k_m (\mathcal{B}_8^A)^{jlm}+ c^A_8(\mathcal{B}_{c\overline 3})_{ij} H(15)^i_{kl}M^k_m (\mathcal{B}_8^A)^{jml} + c^A_9(\mathcal{B}_{c\overline 3})_{ij} H(15)^i_{kl}M^k_m (\mathcal{B}_8^A)^{lmj} \nonumber\\
 &+ c^A_{10}(\mathcal{B}_{c\overline 3})_{ij} H(\overline 6)^i_{kl}M^k_m (\mathcal{B}_8^A)^{jlm}  + c^A_{11}(\mathcal{B}_{c\overline 3})_{ij} H(\overline 6)^i_{kl}M^k_m (\mathcal{B}_8^A)^{jml} + c^A_{12}(\mathcal{B}_{c\overline 3})_{ij} H(\overline 6)^i_{kl}M^k_m (\mathcal{B}_8^A)^{lmj} \nonumber\\
&+ c^A_{13}(\mathcal{B}_{c\overline 3})_{ij} H(15)^m_{kl}M^l_m (\mathcal{B}_8^A)^{ikj}+ c^A_{14}(\mathcal{B}_{c\overline 3})_{ij} H(\overline 6)^m_{kl}M^l_m (\mathcal{B}_8^A)^{ikj} + c^A_{15}(\mathcal{B}_{c\overline 3})_{ij} H(15)^i_{kl}M^m_m (\mathcal{B}_8^A)^{jkl} \nonumber\\
 &+ c^A_{16}(\mathcal{B}_{c\overline 3})_{ij} H(\overline 6)^i_{kl}M^m_m (\mathcal{B}_8^A)^{jkl} + c^A_{17}(\mathcal{B}_{c\overline 3})_{ij} H(\overline 6)^i_{kl}M^m_m (\mathcal{B}_8^A)^{klj}  + c^A_{18}(\mathcal{B}_{c\overline 3})_{ij} H(3_t)_{k}M^i_m (\mathcal{B}_8^A)^{jkm}\nonumber\\
 & + c^A_{19}(\mathcal{B}_{c\overline 3})_{ij} H(3_t)_{k}M^i_m (\mathcal{B}_8^A)^{jmk} + c^A_{20}(\mathcal{B}_{c\overline 3})_{ij} H(3_t)_{k}M^i_m (\mathcal{B}_8^A)^{kmj} + c^A_{21}(\mathcal{B}_{c\overline 3})_{ij} H(3_t)_{k}M^k_m (\mathcal{B}_8^A)^{imj}\nonumber\\
& + c^A_{22}(\mathcal{B}_{c\overline 3})_{ij} H(3_t)_{k}M^m_m (\mathcal{B}_8^A)^{ikj} + c^A_{23}(\mathcal{B}_{c\overline 3})_{ij} H( 3_p)_{k}M^i_m (\mathcal{B}_8^A)^{jkm}+ c^A_{24}(\mathcal{B}_{c\overline 3})_{ij} H( 3_p)_{k}M^i_m (\mathcal{B}_8^A)^{jmk}\nonumber\\
& + c^A_{25}(\mathcal{B}_{c\overline 3})_{ij} H( 3_p)_{k}M^i_m (\mathcal{B}_8^A)^{kmj} + c^A_{26}(\mathcal{B}_{c\overline 3})_{ij} H( 3_p)_{k}M^k_m (\mathcal{B}_8^A)^{imj}+ c^A_{27}(\mathcal{B}_{c\overline 3})_{ij} H( 3_p)_{k}M^m_m (\mathcal{B}_8^A)^{ikj}\nonumber\\
&+ c^A_{28}(\mathcal{B}_{c\overline 3})_{ij} H(15)^m_{kl}M^k_m (\mathcal{B}_8^A)^{ijl} + c^A_{29}(\mathcal{B}_{c\overline 3})_{ij} H(\overline 6)^m_{kl}M^k_m (\mathcal{B}_8^A)^{ijl} + c^A_{30}(\mathcal{B}_{c\overline 3})_{ij} H(3_t)_{k}M^k_m (\mathcal{B}_8^A)^{ijm} \nonumber\\
&+ c^A_{31}(\mathcal{B}_{c\overline 3})_{ij} H(3_t)_{k}M^m_m (\mathcal{B}_8^A)^{ijk}+ c^A_{32}(\mathcal{B}_{c\overline 3})_{ij} H(3_p)_{k}M^k_m (\mathcal{B}_8^A)^{ijm}+ c^A_{33}(\mathcal{B}_{c\overline 3})_{ij} H(3_p)_{k}M^m_m (\mathcal{B}_8^A)^{ijk}\nonumber\\& +c^{c,A}_{1}(\mathcal{B}_{b\overline 3})_{ij} H(3)^c_{k}M^i_l ( \mathcal{B}_8^A)^{jkl}+ c^{c,A}_{2}(\mathcal{B}_{b\overline 3})_{ij} H(3)^c_{k}M^i_l ( \mathcal{B}_8^A)^{jlk}+ c^{c,A}_{3}(\mathcal{B}_{b\overline 3})_{ij} H(3)^c_{k}M^i_l (\mathcal{B}_8^A)^{klj} \nonumber\\
& + c^{c,A}_{4}(\mathcal{B}_{b\overline 3})_{ij} H(3)^c_{k}M^k_l ( \mathcal{B}_8^A)^{ilj}+ c^{c,A}_{5}(\mathcal{B}_{b\overline 3})_{ij} H(3)^c_{k}M^l_l ( \mathcal{B}_8^A)^{ikj}+ c^{c,A}_{6}(\mathcal{B}_{b\overline 3})_{ij} H(3)^c_{k}M^k_l ( \mathcal{B}_8^A)^{ijl}\nonumber\\
& + c^{c,A}_{7}(\mathcal{B}_{b\overline 3})_{ij} H(3)^c_{k}M^l_l ( \mathcal{B}_8^A)^{ijk}.
\end{align}
The relations of the $SU(3)$ irreducible amplitudes and the topological amplitudes are derived as
\begin{align}\label{sol7}
   & c^A_1 = \frac{a^A_1+a^A_2}{8},\quad c^A_2 = \frac{a^A_1-a^A_2}{4},  \quad c^A_3 = \frac{a^A_3}{4},  \quad c^A_4 = \frac{a^A_4}{4}, \quad c^A_5 = \frac{a^A_5+a^A_6}{8}, \quad  c^A_6 = \frac{a^A_5-a^A_6}{4},\quad c^A_7 = \frac{a^A_7+a^A_{10}}{8},\nonumber\\
   &  c^A_{10} = \frac{a^A_7-a^A_{10}}{4},  \quad c^A_8 = \frac{a^A_8 + a^A_{11}}{8},  \quad c^A_{11} = \frac{a^A_8 - a^A_{11}}{4}, \quad c^A_9 = \frac{a^A_9+a^A_{12}}{8}, \quad  c^A_{12} = \frac{a^A_9-a^A_{12}}{4},\quad c^A_{13} = \frac{a^A_{13}+a^A_{14}}{8},\nonumber\\
   &  c^A_{14} = \frac{a^A_{13}-a^A_{14}}{4},  \quad c^A_{15} = \frac{a^A_{15} + a^A_{16}}{8},  \quad c^A_{16} = \frac{a^A_{15} - a^A_{16}}{4}, \quad c^A_{17} = \frac{a^A_{17}}{4},\quad c^A_{28} = \frac{a^A_{28} + a^A_{29}}{8},  \quad c^A_{29} = \frac{a^A_{28} - a^A_{29}}{4},\nonumber\\
   &  c^A_{18} = \frac{3}{8}a^A_1- \frac{1}{8}a^A_2 + \frac{1}{2}a^A_4- \frac{1}{8}a^A_7 +\frac{3}{8}a^A_{10}+ J^S_{18},\quad c^A_{23} = -\frac{1}{8}a^A_1+ \frac{3}{8}a^A_2 - \frac{1}{2}a^A_4+ \frac{3}{8}a^A_7 -\frac{1}{8}a^A_{10}+ a^A_{23},\nonumber\\
   &  c^A_{19} = -\frac{1}{8}a^A_1+ \frac{3}{8}a^A_2 + \frac{1}{8}a^A_{5}- \frac{3}{8}a^A_{6} -\frac{1}{8}a^A_{8}+ \frac{3}{8}a^A_{11}+ a^A_{19},\quad c^A_{24} = \frac{3}{8}a^A_1- \frac{1}{8}a^A_2 - \frac{3}{8}a^A_{5}+ \frac{1}{8}a^A_{6} +\frac{3}{8}a^A_{8}- \frac{1}{8}a^A_{11}+ a^A_{24},\nonumber\\
   &  c^A_{20} = \frac{1}{2}a^A_3- \frac{3}{8}a^A_5 + \frac{1}{8}a^A_{6}- \frac{1}{8}a^A_{9} +\frac{3}{8}a^A_{12}+ a^A_{20},\quad c^A_{25} = -\frac{1}{2}a^A_3+ \frac{1}{8}a^A_5 - \frac{3}{8}a^A_{6}+ \frac{3}{8}a^A_{9} -\frac{1}{8}a^A_{12}+ a^S_{25},
\nonumber\\
   &  c^A_{21} = -\frac{3}{8}a^A_8+ \frac{1}{8}a^A_{11} + \frac{3}{8}a^A_{9}- \frac{1}{8}a^A_{12} -\frac{1}{8}a^A_{13} + \frac{3}{8}a^A_{14}+a^A_{21},\quad  c^A_{26} = \frac{1}{8}a^A_8- \frac{3}{8}a^A_{11} - \frac{1}{8}a^A_{9}+\frac{3}{8}a^A_{12} +\frac{3}{8}a^A_{13} - \frac{1}{8}a^A_{14}+a^A_{26},
\nonumber\\
   &  c^A_{22} = \frac{3}{8}a^A_{13}- \frac{1}{8}a^A_{14} -\frac{3}{8}a^A_{15}+ \frac{1}{8}a^A_{16} -\frac{1}{2}a^A_{17} +a^A_{22},\quad  c^A_{27} = -\frac{1}{8}a^A_{13}+\frac{3}{8}a^A_{14} +\frac{1}{8}a^A_{15}- \frac{3}{8}a^A_{16} +\frac{1}{2}a^A_{17} +a^A_{27},
   \nonumber\\
   &  c^A_{30} = -\frac{3}{8}a^A_{7}+ \frac{1}{8}a^A_{10} +\frac{3}{8}a^A_{28}- \frac{1}{8}a^A_{29} +a^A_{30},\quad  c^A_{32} = \frac{1}{8}a^A_{7}- \frac{3}{8}a^A_{10} -\frac{1}{8}a^A_{28}+ \frac{3}{8}a^A_{29} +a^A_{32},   \nonumber\\
   &  c^A_{31} = \frac{1}{8}a^A_{15}- \frac{3}{8}a^A_{16} -\frac{1}{8}a^A_{28}+ \frac{3}{8}a^A_{29} +a^A_{31},\quad  c^A_{33} = -\frac{3}{8}a^A_{15}+ \frac{1}{8}a^A_{16} +\frac{3}{8}a^A_{28}- \frac{1}{8}a^A_{29} +a^A_{33},\nonumber\\
& c^{c,A}_1 =  a^{c,A}_1,\qquad c^{c,A}_2 =  a^{c,A}_2,\qquad c^{c,A}_3 =  a^{c,A}_3,\qquad c^{c,A}_4 =  a^{c,A}_4,\qquad  c^{c,A}_5 =  a^{c,A}_5,\nonumber\\ & c^{c,A}_6 =  a^{c,A}_6,\qquad  c^{c,A}_7 =  a^{c,A}_7.
\end{align}
By substituting Eqs.~\eqref{sy1} and \eqref{sy2} into each term of Eqs.~\eqref{IS} and \eqref{IA}, the relations between the $SU(3)$ irreducible amplitudes constructed by $3$-rank and $(1,1)$-rank octets are derived as
\begin{align}\label{sol10}
& b_1 = -(2c^A_4+c^A_{10}-c^A_{12})/\sqrt{2}
   +(2c^S_6-c^S_{10}+2c^S_{11}-c^S_{12})/\sqrt{6},\nonumber\\
& b_2 = (c^A_2-2c^A_{3}+c^A_{6})/\sqrt{2}
   +(3c^S_2-c^S_{6})/\sqrt{6},\nonumber\\
& b_3 = -(c^A_6-c^A_{16}+2c^A_{17})/\sqrt{2}
   +(c^S_6+3c^S_{16})/\sqrt{6},\nonumber\\
& b_4 = (c^A_6+c^A_{11}+c^A_{12})/\sqrt{2}
   -(c^S_6+2c^S_{10}-c^S_{11}-c^S_{12})/\sqrt{6},\nonumber\\
& b_5 = (c^A_2-2c^A_{3}+c^A_{6}+c^A_{11}+c^A_{12}+c^A_{14}-2c^A_{29})/\sqrt{2}
   +(3c^S_2-c^S_{6}-2c^S_{10}+c^S_{11}+c^S_{12}-3c^S_{14})/\sqrt{6},\nonumber\\
& b_6 = (c^A_7-c^A_{9})/\sqrt{2}
   +(c^S_7-2c^S_{8}+c^S_{9})/\sqrt{6},\nonumber\\
& b_7 = (c^A_1+c^A_{5})/\sqrt{2}
   -(c^S_1-2c^S_{3}+2c^S_4-c^S_{5})/\sqrt{6},\nonumber\\
& b_8 = -(c^A_5-c^A_{15})/\sqrt{2}
   +(2c^S_4-c^S_{5}-c^S_{15}+2c^S_{17})/\sqrt{6},\nonumber\\
& b_9 = (c^A_5+c^A_{8}+c^A_{9})/\sqrt{2}
   -(2c^S_4-c^S_{5}+2c^S_7-c^S_{8}-c^S_{9})/\sqrt{6},\nonumber\\
& b_{10} = -(c^A_1+c^A_{5}+c^A_{8}+c^A_{9}-c^A_{13}-2c^A_{28})/\sqrt{2}
   +(c^S_1-2c^S_{3}+2c^S_{4}-c^S_{5}+2c^S_{7}
   -c^S_{8}-c^S_{9}-3c^S_{13})/\sqrt{6},\nonumber\\
& b_{11} = (c^A_{23}-c^A_{25})/\sqrt{2}
   +(c^S_{23}-2c^S_{24}+c^S_{25})/\sqrt{6},\nonumber\\
& b_{12} = -(c^A_{24}+c^A_{25}-c^A_{27}-2c^A_{33})/\sqrt{2}
   +(2c^S_{23}-c^S_{24}-c^S_{25}-3c^S_{27})/\sqrt{6},\nonumber\\
& b_{13} = (c^A_{24}+c^A_{25})/\sqrt{2}
   -(2c^S_{23}-c^S_{24}-c^S_{25})/\sqrt{6},\nonumber\\
& b_{14} = -(c^A_{23}-c^A_{25}-c^A_{26}-2c^A_{32})/\sqrt{2}
   -(c^S_{23}-2c^S_{24}+c^S_{25}+3c^S_{26})/\sqrt{6},\nonumber\\
& b_{15} = (c^A_{18}-c^A_{20})/\sqrt{2}
   +(c^S_{18}-2c^S_{19}+c^S_{20})/\sqrt{6},\nonumber\\
& b_{16} = -(c^A_{19}+c^A_{20}-c^A_{22}-2c^A_{31})/\sqrt{2}
   +(2c^S_{18}-c^S_{19}-c^S_{20}-3c^S_{22})/\sqrt{6},\nonumber\\
& b_{17} = (c^A_{19}+c^A_{20})/\sqrt{2}
   -(2c^S_{18}-c^S_{19}-c^S_{20})/\sqrt{6},\nonumber\\
& b_{18} = -(c^A_{18}-c^A_{20}-c^A_{21}-2c^A_{30})/\sqrt{2}
   -(c^S_{18}-2c^S_{19}+c^S_{20}+3c^S_{21})/\sqrt{6},
\nonumber\\
& b^c_{1}  = (-c^{c,A}_{1}+c^{c,A}_{3}+c^{c,A}_{4}+2c^{c,A}_{6})/\sqrt{2}-(
 c^{c,S}_{1}-2c^{c,S}_{2}
 +c^{c,S}_{3}+3c^{c,S}_{4})/\sqrt{6}, \nonumber\\
& b^c_{2}  = (-c^{c,A}_{2}-c^{c,A}_{3}+c^{c,A}_{5}+2c^{c,A}_{7})/\sqrt{2}-(
 -2c^{c,S}_{1}+c^{c,S}_{2}
 +c^{c,S}_{3}+3c^{c,S}_{5})/\sqrt{6}, \nonumber\\
& b^c_{3}  = (c^{c,A}_{2}+c^{c,A}_{3})/\sqrt{2}-(2c^{c,S}_{1}-c^{c,S}_{2}
 -c^{c,S}_{3})/\sqrt{6}, \nonumber\\
& b^c_{4}  = (c^{c,A}_{1}-c^{c,A}_{3})/\sqrt{2}-(-c^{c,S}_{1}+2c^{c,S}_{2}
 -c^{c,S}_{3})/\sqrt{6}.
\end{align}
By substituting Eqs.~\eqref{sol6} and \eqref{sol7} into Eq.~\eqref{sol10}, Eq.~\eqref{sol11} is re-derived.

As noted in Ref.~\cite{Wang:2024ztg}, the amplitudes constructed by 3-rank or $(1,1)$-rank octets can be decomposed by the other amplitudes constructed by 3-rank or $(1,1)$-rank octets, and the inverse solutions exist.
The amplitudes constructed by $(1,1)$-rank octets can be decomposed by the amplitudes constructed by 3-rank octets. However, the amplitudes constructed by 3-rank octets cannot be decomposed by the amplitudes constructed by $(1,1)$-rank octets.

\end{appendix}

%%%%%%%%%%%%%%%%%%%%%%%%%%%%%%%%%%%%%%%%%%%%%%%%%%%%%%%%%%%%%%%
%\include{reference}

\end{document}